\documentclass[a4paper,fleqn,usenatbib]{mnras}

\usepackage[T1]{fontenc}
\usepackage{ae,aecompl}

\usepackage{graphicx}	
\usepackage{amsmath}	
\usepackage{amssymb}	

\title[Parameter constraints in Modeling of 3C 279]{Parameter constraints in a near-equipartition model with multi-frequency \emph{NuSTAR}, \emph{Swift} and \emph{Fermi-LAT} data from 3C 279}
\author[D. H. Yan, L. Zhang, S. N. Zhang]{Dahai Yan$^{1}$\thanks{E--mail: yandahai@ihep.ac.cn}, Li Zhang$^{2}$\thanks{E--mail: lizhang@ynu.edu.cn}, Shuang-Nan Zhang$^1$\thanks{E--mail: zhangsn@ihep.ac.cn}\\
$^1$Key Laboratory of Particle Astrophysics, Institute of High Energy Physics,
Chinese Academy of Sciences, Beijing 100049, China\\
$^2$Department of Astronomy, Yunnan University, Key Laboratory of Astroparticle Physics of Yunnan Province, Kunming, 650091, China}

\date{Accepted XXX. Received YYY; in original form ZZZ}

\pubyear{2015}

\begin{document}
\label{firstpage}
\pagerange{\pageref{firstpage}--\pageref{lastpage}}
\maketitle

\begin{abstract}
Precise spectra of 3C 279 in the 0.5-70 keV range, obtained during two epochs of
 \emph{Swift} and \emph{NuSTAR} observations, are analyzed using a near-equipartition model.
We apply a one-zone leptonic model with a three-parameter log-parabola electron energy distribution (EED) to fit the \emph{Swift} and \emph{NuSTAR} X-ray data, as well as simultaneous optical and \emph{Fermi}-LAT $\gamma$-ray data.
The Markov Chain Monte Carlo (MCMC) technique is used to search the high-dimensional parameter space and evaluate the uncertainties on model parameters. We show that the two spectra can be successfully fit in near-equipartition conditions, defined by the ratio of the energy density of relativistic electrons to magnetic field $\zeta_{\rm e}$ being close to unity. In both spectra, the observed X-rays are dominated by synchrotron-self Compton photons, and the observed $\gamma$ rays are dominated
by  Compton scattering of external infrared photons from a surrounding dusty torus.
 Model parameters are well constrained. From the low state to the high state, both the curvature of the log-parabola width parameter and the synchrotron peak frequency significantly increase. The derived magnetic fields in the two states are nearly identical ($\sim1$\ G), but the Doppler factor in the high state is larger than that in the low state ($\sim$28 versus $\sim$18). We derive that the gamma-ray emission site takes place outside the broad-line region, at $\gtrsim$ 0.1 pc from the black hole, but within the dusty torus. Implications for 3C 279 as a source of high-energy cosmic-rays are discussed.
\end{abstract}


\section{Introduction}
Blazar emission is generally dominated by non-thermal radiation
over all frequencies ranging from radio to TeV $\gamma$ rays.
By modeling blazar spectra,
emission mechanisms and physical properties of the relativistic jets can be determined.
The typical multi-wavelength spectral energy distribution (SED) of a
blazar is characterized by two distinct humps.
It is generally accepted that the first hump in the blazar SED is
nonthermal synchrotron emission radiated by relativistic electrons in the jet.
The origin of the emission in the
$\gamma$-ray hump is less certain. Leptonic and hadronic models
have both been proposed to explain the
formation of the second bump \citep[see, e.g.,][for review]{bhk12}.
In general, both types of models are able to reproduce the SEDs well,
but require quite different jet properties, with the
hadronic models sometimes requiring super-Eddington
jet powers \citep{bottcher13,Zdziarski}.
The leptonic models are more attractive by allowing
much smaller jet powers, and by more readily
explaining  rapid blazar variability
from highly radiative electrons in contrast to the more
weakly radiating hadrons \cite[cf.][]{2015MNRAS.448..910C}.

All hadronic models are hybrid lepto-hadronic, because the synchrotron
bump is believed to be radiated by directly accelerated electron/lepton
primaries in the jet. Besides the leptonic synchrotron self-Compton (SSC) component,
$\gamma$ rays in hadronic models can result from
proton- or ion-synchrotron radiation \citep{Aharonian2000,mucke2003} and $p\gamma$
interactions, including both photopion and Bethe-Heitler
photopair production  \citep{ad01,ad03,bottcher09,bottcher13,murase14,masti13,dimit2014,Weidinger2015,Yan15}.
Interactions of ultra-relativistic protons and ions in the jet with photons
induce cascades by the injection of extremely high-energy photons and leptons.

In leptonic models, $\gamma$ rays are produced via inverse Compton (IC) scattering of the relativistic electrons, including SSC scattering  \citep[e.g.,][]{Maraschi1992,Tavecchio98,Finke08,Yan14} and external Compton (EC) scattering \citep[e.g.,][]{Dermer93,Sikora1994,bl,Dermer09,Kang,Paliya}.
EC models seem to be required
 for flat-spectrum radio quasars (FSRQs; a subclass of blazars with intense broad-line radiation in the inner-jet).
Depending on the emission site, the external seed photons could be dominated by radiation from accretion-disk \citep[e.g.,][]{Dermer92,Dermer93,Dermer02}, broad-line region (BLR) \citep[e.g.,][]{Sikora1994,zhang12}, and dusty torus \citep[e.g.,][]{bl}.
Many efforts have been made to locate the $\gamma$-ray emission regions of FSRQs  \citep[e.g.,][]{liu06,bai,Tavecchio2010,Yan12,Dotson,Nalewajko14},
but it is still an unresolved issue.

Radio telescopes can partially resolve the structure of blazar jets.
The observed connections between $\gamma$-ray flares and radio/millimeter flares are used to
argue for gamma-ray emission arising at distance
scales of $\gtrsim\ $pc, or even $\gtrsim10\ $pc \citep[e.g.,][]{Jorstad01,Jorstad10,Agudo}.
However, \citet{Tavecchio2010} argued that the rapid variabilities on timescale of a few hours challenge the scenario
in which $\gamma$-ray emission site is placed at tens of parsec from the black hole.
Using the time lags of $\gamma$ rays relative to broad emission lines, \citet{liu11} argued that $\gamma$-ray emission may be produced in the regions of the jet at $\lesssim2\ $pc from the black hole.

Although blazar emission models have been constructed
that accurately reproduce contemporaneous SED
data, the number of free parameters in even the simplest SSC model
can exceed the number of constraining observables.
Therefore, blazar SED modeling generally yields degenerate parameter
values, which makes it difficult to investigate
correlations between model parameters.

Claims and explanations of correlations between
model parameters have been made,
for example, for statistical quantities
like the blazar sequence \citep{fos98,ghi98},
or for flux-flux or flux-index correlations
between different spectral states of the same
blazar. Due to many free and unconstrained parameters,
it is difficult to establish robust correlations between model
parameters using simple fitting methods.

Using the Markov Chain Monte Carlo (MCMC) technique,
\citet{Yan13} constrained parameter space in an SSC model for Mrk 421.
The high-quality SEDs enable us to
derive the confidence intervals of each model parameter using the MCMC method, and
to do a comparison of model fits for different SEDs.
In a series of papers \citep{Yan13,Peng,Zhou14}, we
showed that fitting high-quality SEDs with the MCMC technique
is a powerful approach to investigating the blazar jet physics.
In a related modeling effort,
\citet{Dermer14} assumed that a three-parameter log-parabola
function approximates the electron energy distribution (EED).
Introducing the equipartition parameter $\zeta_e = u_e^\prime/u_{B^\prime}^\prime$  as the ratio of
the energy densities of electrons $u_e^\prime$ and magnetic field $u^\prime_{B^\prime}$ gives, when $\zeta_e\approx 1$, the minimum synchrotron jet power.
Besides $\zeta_e$, the other three equipartition factors in this modeling approach \citep{Cerruti,Dermer14} are:
(1) $\zeta_{\rm s}$, the ratio of the energy density of synchrotron radiation $u^{\prime}_{\rm syn}$ to $u^\prime_{B^\prime}$;
(2) $\zeta_{\rm BLR}$, the ratio of the energy density of BLR radiation in comoving frame $u^{\prime}_{\rm BLR}$ to $u^\prime_{B^\prime}$; and
(3) $\zeta_{\rm dust}$, the ratio of the energy density of dust radiation in comoving frame $u^{\prime}_{\rm dust}$ to $u^\prime_{B^\prime}$.
In this approach, the physical parameters (e.g., fluid magnetic field $B^\prime$
and Doppler factor $\delta_{\rm D}$) are expressed in terms of observables (e.g., synchrotron peak frequency and  synchrotron peak luminosity, and variability time scale).
With this formulation,
GeV cutoffs in blazar spectra are explained under equipartition conditions by IC scattering broad-line region (BLR) radiation in Klein-Nihisna regime \citep{Cerruti}.
In SSC models, the equipartition modeling approach allows
the complete set of parameters to be fully constrained, contingent on the underlying
assumption that the electron distribution can be described, at least approximately,
by a log-parabola function. Here, we constrain the parameters using the
near-equipartition, log-parabola model for 3C 279.
In our approach, we do not adopt $\zeta_{\rm BLR}$ and $\zeta_{\rm dust}$ to define the energy densities of the external radiations. Instead, we use the distance $r$ from emitting region to super-massive black hole to deduce the energy densities of the external radiations via the standard scaling relations for BLR and dust torus radiations. We will give the details in Section 2.

The blazar 3C 279, at redshift $z=0.536$, is classified as an FSRQ
due to its strong prominent emission lines, and was one of the
first $\gamma$-ray blazars discovered \citep{1992ApJ...385L...1H}.
Its synchrotron peak frequency is $< 10^{14}$\ Hz, making it a low-synchrotron
peaked object, like most FSRQs \citep{2011ApJ...743..171A}.
Rapid variability is observed in 3C 279, the analysis in \citet{Paliya2} shows that the variability timescale for GeV emission of 3C 279 can be down to $\sim$1 - 2 hour.

Using the near-equipartition
leptonic model with a log-parabola electron energy
distribution (EED), \cite{Dermer14} satisfactorily
explained four SEDs of 3C 279
reported in \citet{Hayashida}.
Since then, detailed simultaneous multiwavelength
SED data of 3C 279 in 4 states of elevated activity that took place between
2013 December to 2014 April were analyzed  by \citet{Hayashida15}.
In particular, \citet{Hayashida15} reported 0.5-70 keV X-ray spectra obtained from
\emph{Swift}-XRT and \emph{NuSTAR} observations in two flaring periods,
namely Period A, taking place from 16 -- 19 December 2013, and Period C, from 31 December -- 3 January 2014.
\citet{Hayashida15} argue that their
one-zone leptonic models with a double-broken power-law EED
fail to explain the \emph{Swift}-XRT and \emph{NuSTAR} X-ray spectra.

In this paper, we fit the two simultaneous SEDs that include \emph{NuSTAR} data reported in \citet{Hayashida15}
using the near-equipartition blazar model with a log-parabola EED. Using the MCMC technique, we derive the best-fit
results and the uncertainties on parameters.
The implications of the results on the acceleration processes in the jet of 3C 279 are discussed.
Supposing that 3C 279 is a high-energy cosmic-ray (CR) source,
we discuss its implied CR luminosity.
We use parameters $H_0=71\rm \ km\ s^{-1}\ Mpc^{-3}$, $\Omega_{\rm m}=0.27$, and $\Omega_{\Lambda}=0.73$.

\section{Emission model and fitting technique}

In the log-parabola leptonic blazar model \citep{Dermer14},
the non-thermal electron distribution is assumed to be isotropic in the emission region (blob), and
 described by a log-parabola function
\begin{equation}
\gamma'^2 N'_e(\gamma')\sim\left( \frac{\gamma'}{\gamma'_{\rm pk}} \right)^{-b\ \log{(\gamma'/\gamma'_{\rm pk})}}\;,
\end{equation}
where $\gamma'$ is the electron Lorentz factor, $b$ is the spectral curvature parameter, and $\gamma'_{\rm pk}$ is the peak Lorentz factor in the $\gamma^{\prime 2}N'_e(\gamma')$ distribution. Emission from the blob is strongly boosted due to the beaming effect. Besides SSC emission, EC components
for low-energy target photons from the broad-line region (BLR; EC-BLR) and the dusty torus (EC-dust) are included.

As mentioned in Section 1, we do not use the equipartition relations $\zeta_{\rm BLR}$ and $\zeta_{\rm dust}$ \citep{Cerruti,Dermer14} to normalize the energy densities of the external radiations.
We take the distance $r$ from the central black hole to the emitting blob as an input parameter.
The energy densities of BLR radiation ($u_{\rm BLR}$) and dust radiation ($u_{\rm dust}$) can be expressed as functions of $r$ \citep{Sikora09,Hayashida}
\begin{equation}
u_{\rm BLR} (r)=\frac{\tau_{\rm BLR} L_{\rm disk}}{4\pi r^2_{\rm BLR}c[1+(r/r_{\rm BLR})^3]},\
\label{u1}
\end{equation}
\begin{equation}
u_{\rm dust} (r)=\frac{\tau_{\rm dust} L_{\rm disk}}{4\pi r^2_{\rm dust}c[1+(r/r_{\rm dust})^4]}.\
\label{u2}
\end{equation}
The size of BLR is related to the disk luminosity $L_{\rm disk}$: $r_{\rm BLR}=10^{17}(L_{\rm disk}/10^{45}\rm \ erg\ s^{-1})^{1/2}\ $cm \citep{ghisellini09,ghisellini14}. We assume a dust torus with the size of $r_{\rm dust}=10^{18}(L_{\rm disk}/10^{45}\rm \ erg\ s^{-1})^{1/2}\ $cm. This size of dust torus is half of the size used in \cite{ghisellini14}, which enhances the energy density of dust radiation. Using the scalings between $L_{\rm disk}$ and $r_{\rm BLR/dust}$, we can rewrite Eqs.~(\ref{u1}) and (\ref{u2}) as
\begin{equation}
u_{\rm BLR} (r)\simeq\frac{0.3\tau_{\rm BLR}}{1+(r/r_{\rm BLR})^3}\rm \ erg\ cm^{-3},\
\label{u3}
\end{equation}
\begin{equation}
u_{\rm dust} (r)\simeq\frac{0.003\tau_{\rm dust}}{1+(r/r_{\rm dust})^4}\rm \ erg\ cm^{-3}, \
\label{u4}
\end{equation}
where $\tau_{\rm BLR}$ and $\tau_{\rm dust}$ are the fractions of the disk luminosity reprocessed into BLR radiation
and into dust radiation, respectively. We adopt the typical values of $\tau_{\rm BLR}=0.1$ \citep[e.g.,][]{ghisellini14} and $\tau_{\rm dust}=0.3$ \citep[e.g.,][]{Hao,Malmrose}.
Using Eqs. (\ref{u3}) and (\ref{u4}), one can obtain $u_{\rm BLR} (\rm erg\ cm^{-3})\lesssim0.3 \tau_{\rm BLR}$, and $u_{\rm dust} (\rm erg\ cm^{-3})\lesssim3\tau_{\rm dust}\times10^{-3}$. The two energy densities are limited by $r_{\rm BLR}$ and $r_{\rm dust}$, respectively. When $r$ is smaller than $r_{\rm dust}$, $u_{\rm dust}$ varies by a factor of at most 2. Moreover, the energy density of the IR dust radiation should be lower than the energy density of a blackbody with the temperature of $T_{\rm dust}$, namely $u_{\rm dust}<u_{\rm bb} (T_{\rm dust})\backsimeq3\times10^{-4}(T_{\rm dust}/440\rm\ K)^4$.

The external radiation is assumed to be a dilute blackbody radiation, namely a blackbody spectral shape normalized to $u_{\rm BLR/dust}$. BLR radiation is dominated by $\rm Ly\alpha$ line photons.
We adopt an effective temperature for the BLR radiation of $T_{\rm BLR}=6.3\times10^4\ $K,
so that the energy density of BLR radiation peaks at
$\approx 2.82 k_{\rm B} T/h \cong 3.7\times 10^{15}$ Hz
in the $u(\nu)$ distribution \citep{Tavecchio08}.
We consider IR dust radiation with $T_{\rm dust}=800\ $K.
Note that \citet{Dermer14} approximated the BLR and dust photon fields as monochromatic.
\citet{Cerruti} considered a complex of atomic emission lines in BLR.
The different approximations for the BLR and dust radiations do not significantly modify the modeling results.

The input parameters in the model are:
(i) $L_{48}= L_{\rm syn}/10^{48}~ \textrm{erg s}^{-1}$, the apparent isotropic bolometric synchrotron luminosity;
(ii) $\nu_{14}=(1+z){\nu_{\rm syn}^{\rm obs}}/{10^{14}\ \textrm{Hz}}$, the synchrotron peak frequency in the source frame, where $\nu_{\rm syn}^{\rm obs}$ is the measured synchrotron peak frequency;
(iii) $t_4=t_{\rm obs, var}/[(1+z)10^4\ \textrm{s})]$, the source variability timescale, where $t_{\rm obs, var}$ is the minimum measured variability timescale; as already noted, we define the equipartition parameter
(iv) $\zeta_e={u'_{\rm e}}/{u'_B}$;
(v) $\zeta_s= {u'_{\rm syn}}/{u'_B}$, the ratio between the synchrotron photon $(u'_{\rm syn})$ and magnetic-field energy densities;
(vi) $b$, curvature parameters of EED;
(vii) $r$, the location of emitting blob along the jet;
(viii) $T_{\rm dust}$, the temperature of dust torus IR radiation; and
(ix) $L_{\rm disk}$, accretion disk luminosity.

There are six output parameters: (1) Doppler factor $\delta_{\rm D}$; (2) fluid magnetic-field strength $B'$; (3) $\gamma'_{\rm pk}$; (4) $u_{\rm BLR}$; (5) $u_{\rm dust}$; and (6) comoving radius of emitting blob, $R'=c\delta_{\rm D}t_{\rm obs, var}/(1+z)$.
The physical model parameters, $\delta_{\rm D}$, $B'$, and $\gamma'_{\rm pk}$, are deduced by using the equations in \citet{Dermer14,Dermer15}. The SSC and EC emissions are calculated using the methods given in \citet{Dermer09}. Synchrotron self-absorption (SSA) is included. The contribution of thermal emission is included, which is assumed to be from a Shakura-Sunyaev disk, using the expression following eq.\ (9) in  \citet{Dermer14}.

We use the MCMC technique to do the fitting \citep[see details in][]{yuan11,liu12,Yan13}.
The MCMC method, based on the Bayesian statistics,
is a powerful tool for high-dimensional parameter space investigation.
The Metropolis-Hastings sampling algorithm which ensures that the probability
density functions of model parameters can be asymptotically
approached with the number density of samples, is adopted to determine
the jump probability from one point to the next in parameter
space \citep{MCMC0,MCMC}. The MCMC method is also an effective approach for determining uncertainties of model parameters.

\begin{figure}
	   \centering
		\includegraphics[width=270pt,height=210pt]{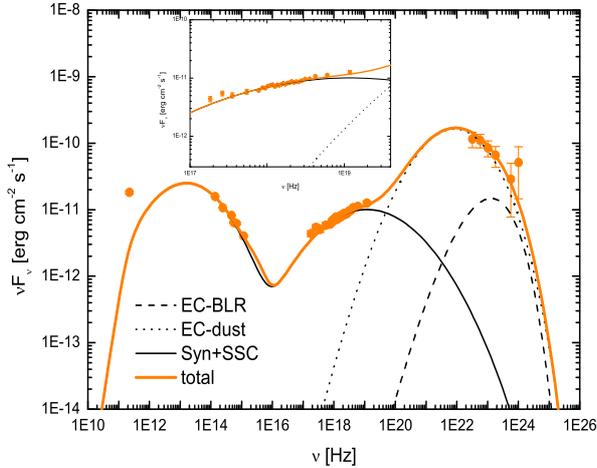}
	  \caption{ Best-fit model to the SED of 3C 279 for Period A \citep[data from][]{Hayashida15}. The inset shows the details of fit at X-ray energies. \label{seda}}
\end{figure}

\begin{figure*}
	   \centering
		\includegraphics[width=240pt,height=180pt]{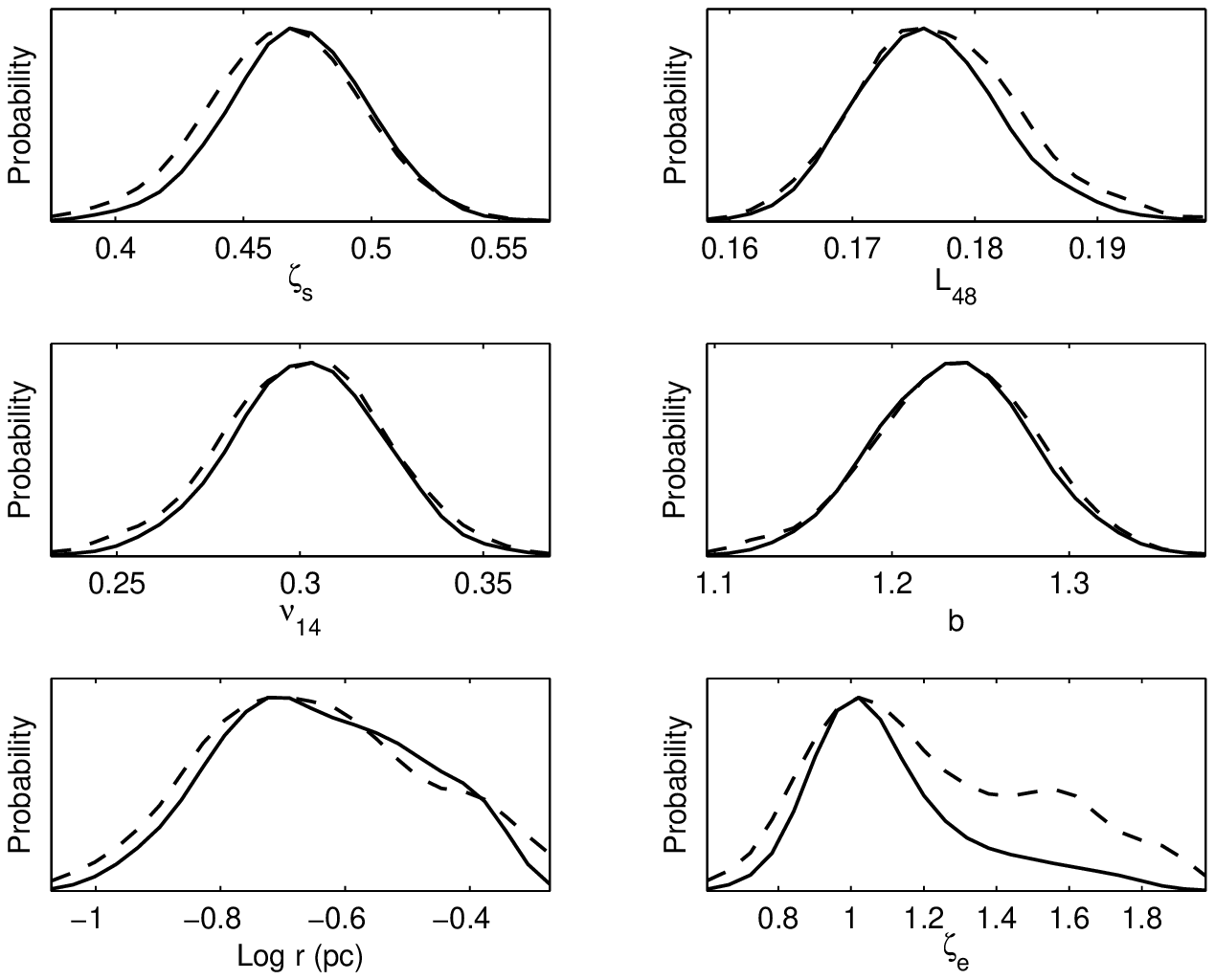}
		\includegraphics[width=240pt,height=180pt]{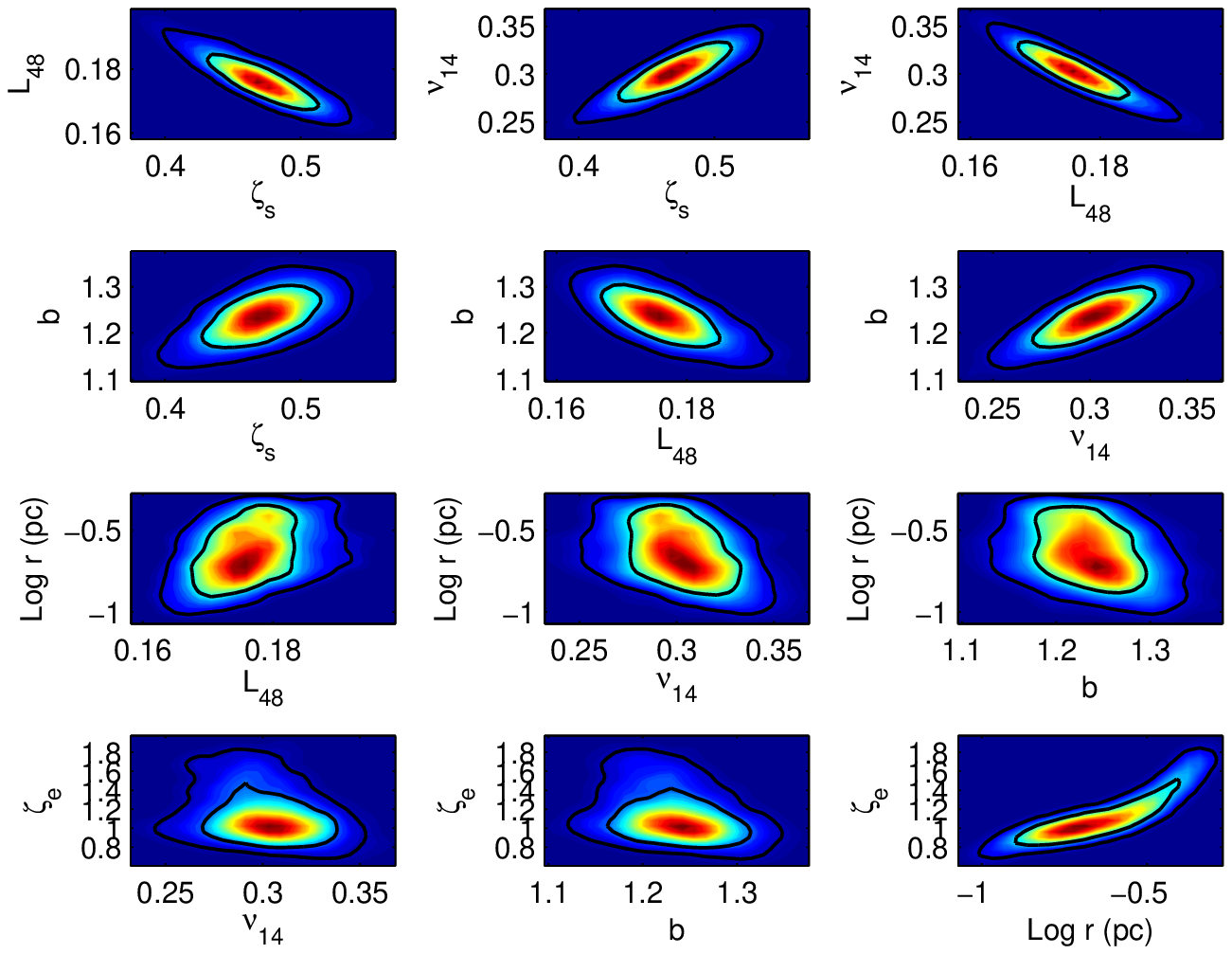}
        \includegraphics[width=240pt,height=180pt]{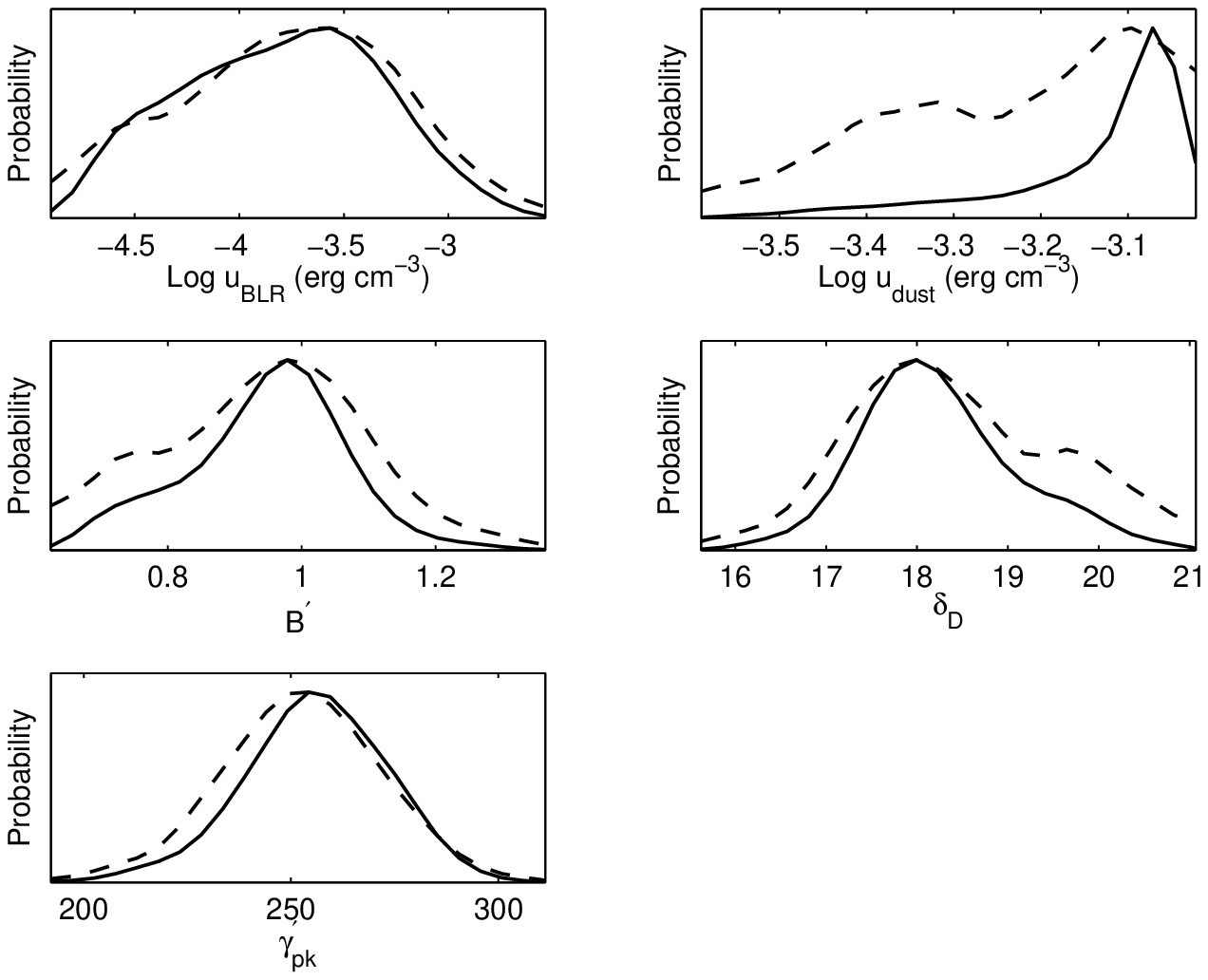}
		\includegraphics[width=240pt,height=180pt]{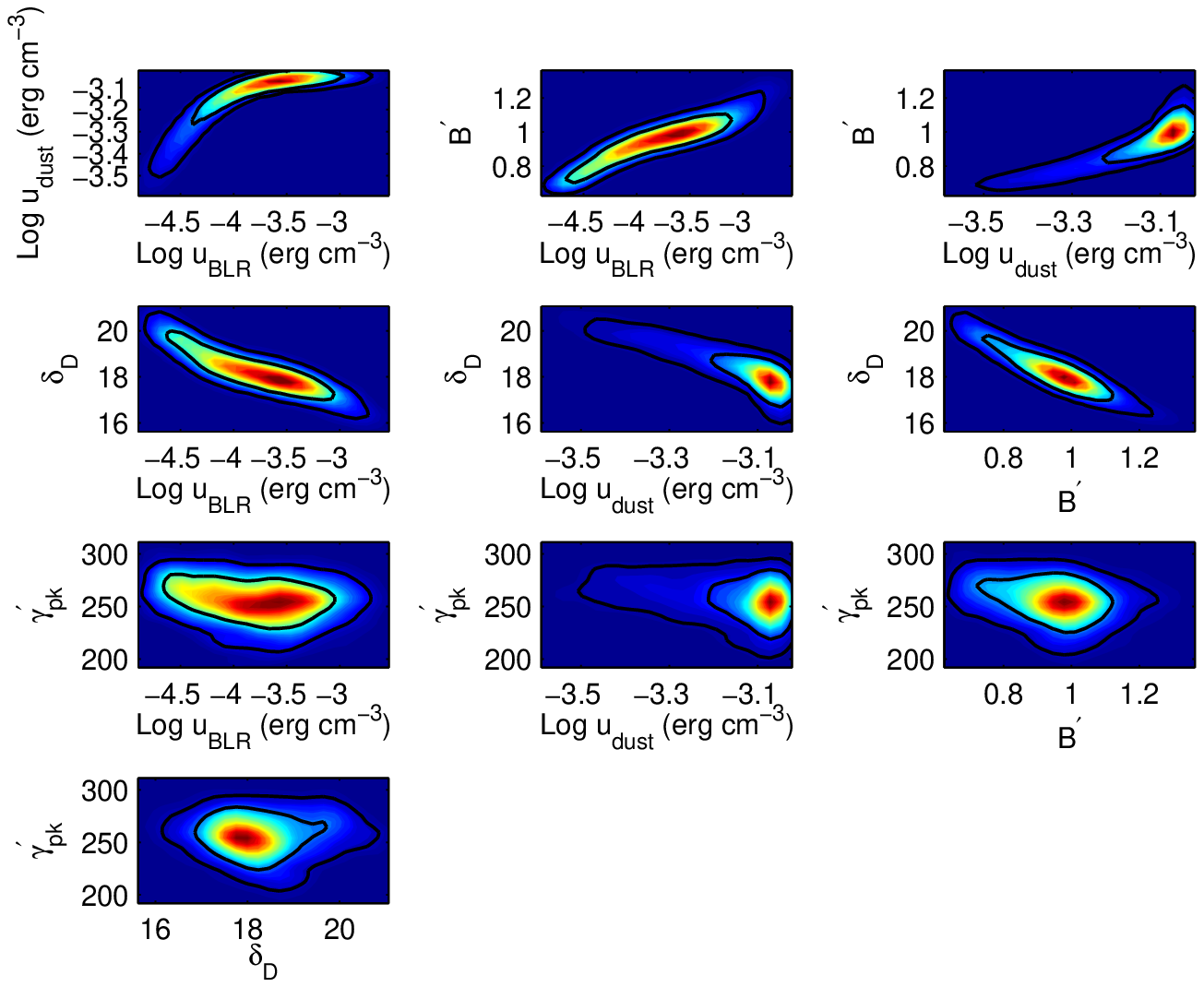}
	  \caption{One-dimensional probability distribution of parameter values (left; the dashed curves are the mean likelihoods of
samples and the solid curves are the marginalized probabilities) and two-dimensional contours of parameters (right; the regions enclosing 68\%(95\%) confidence level are shown) for Period A. The upper panel is for the input parameters, and the lower panel is for the output parameters. \label{dista}}
\end{figure*}

\begin{figure}
	   \centering
		\includegraphics[width=270pt,height=210pt]{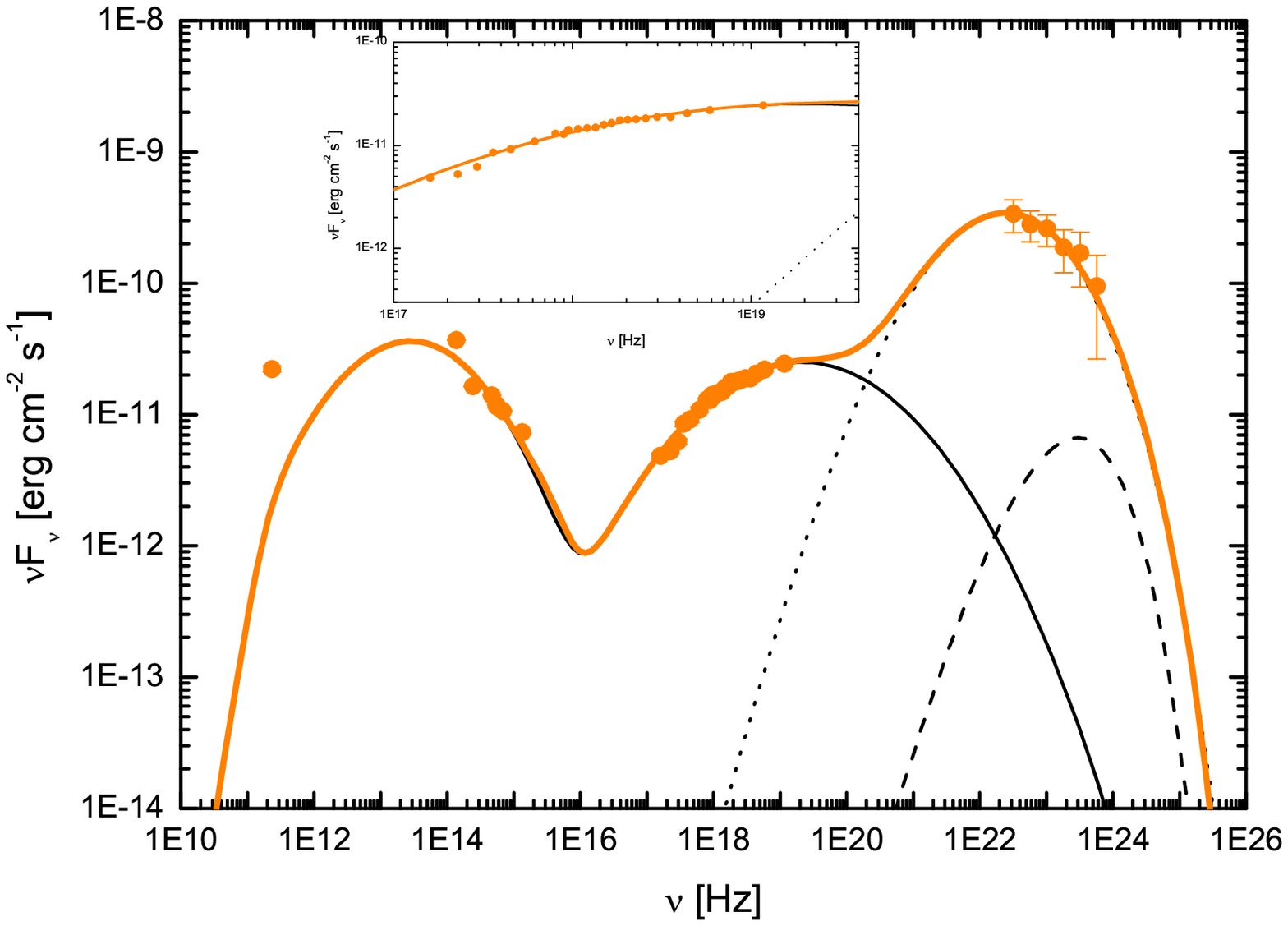}
	  \caption{Best-fit model to the SED of 3C 279 for Period C \citep[data from][]{Hayashida15}. The inset shows the details of fit at X-ray energies. \label{sedc}}
\end{figure}

\begin{figure*}
	   \centering
		\includegraphics[width=240pt,height=180pt]{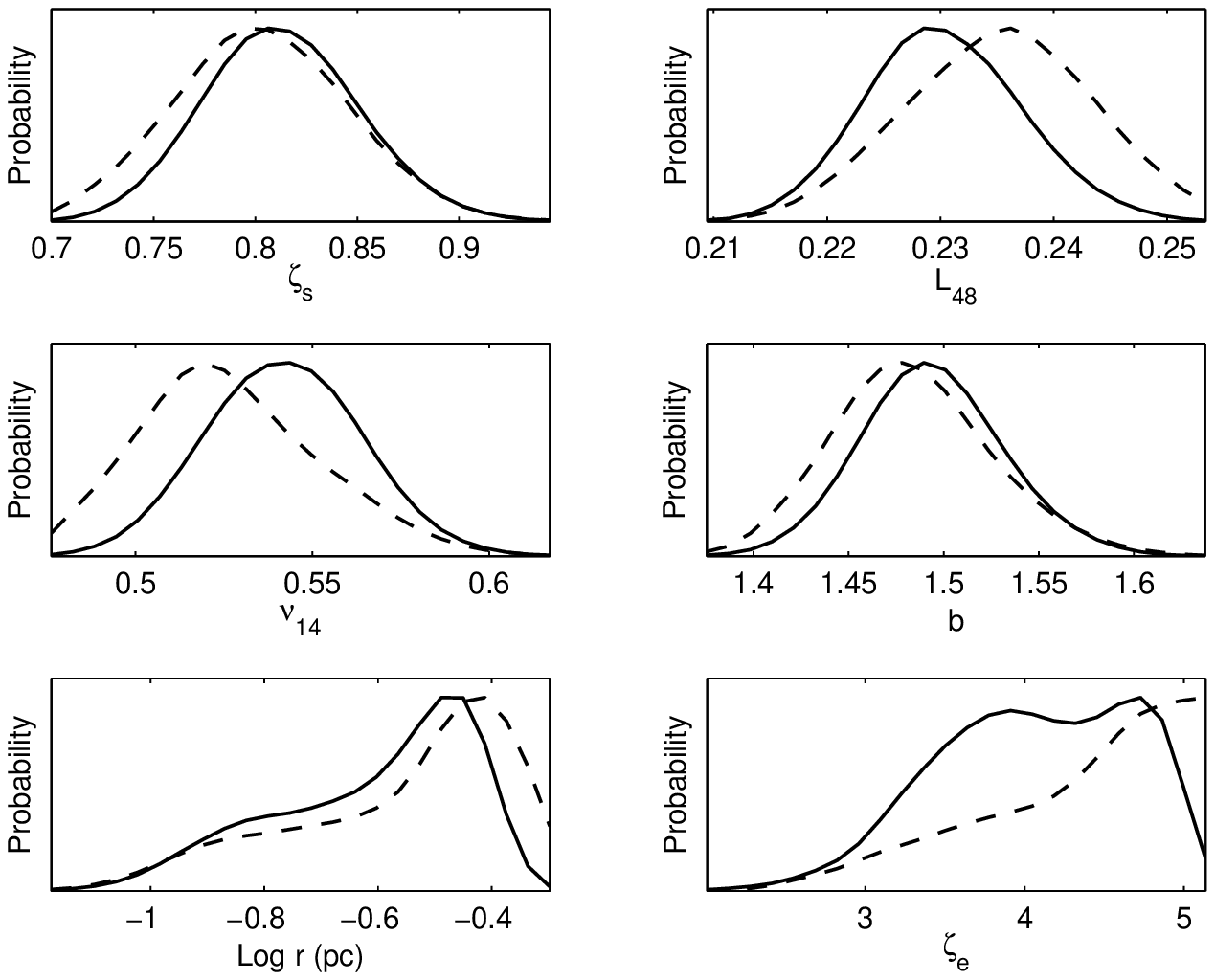}
		\includegraphics[width=240pt,height=180pt]{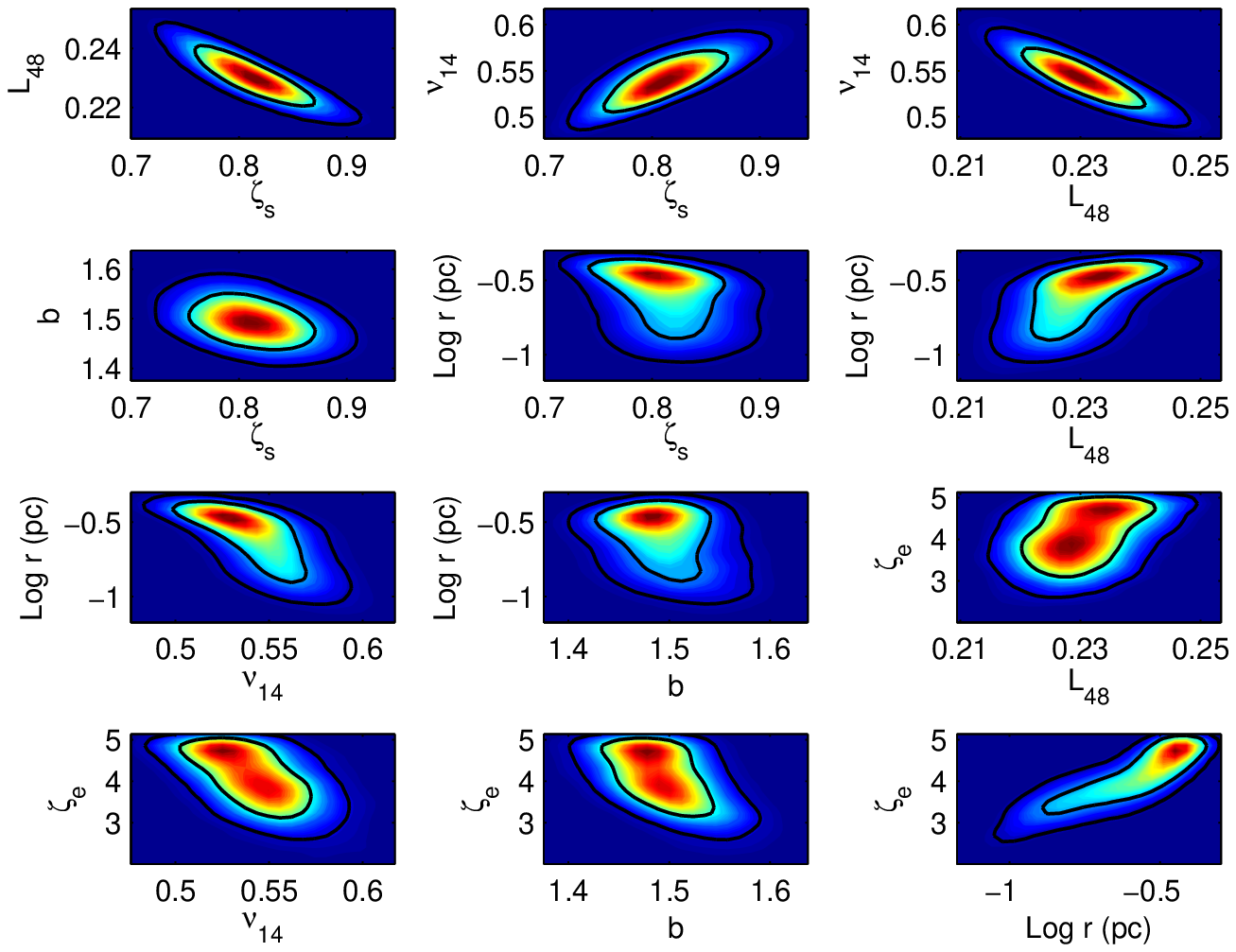}
        \includegraphics[width=240pt,height=180pt]{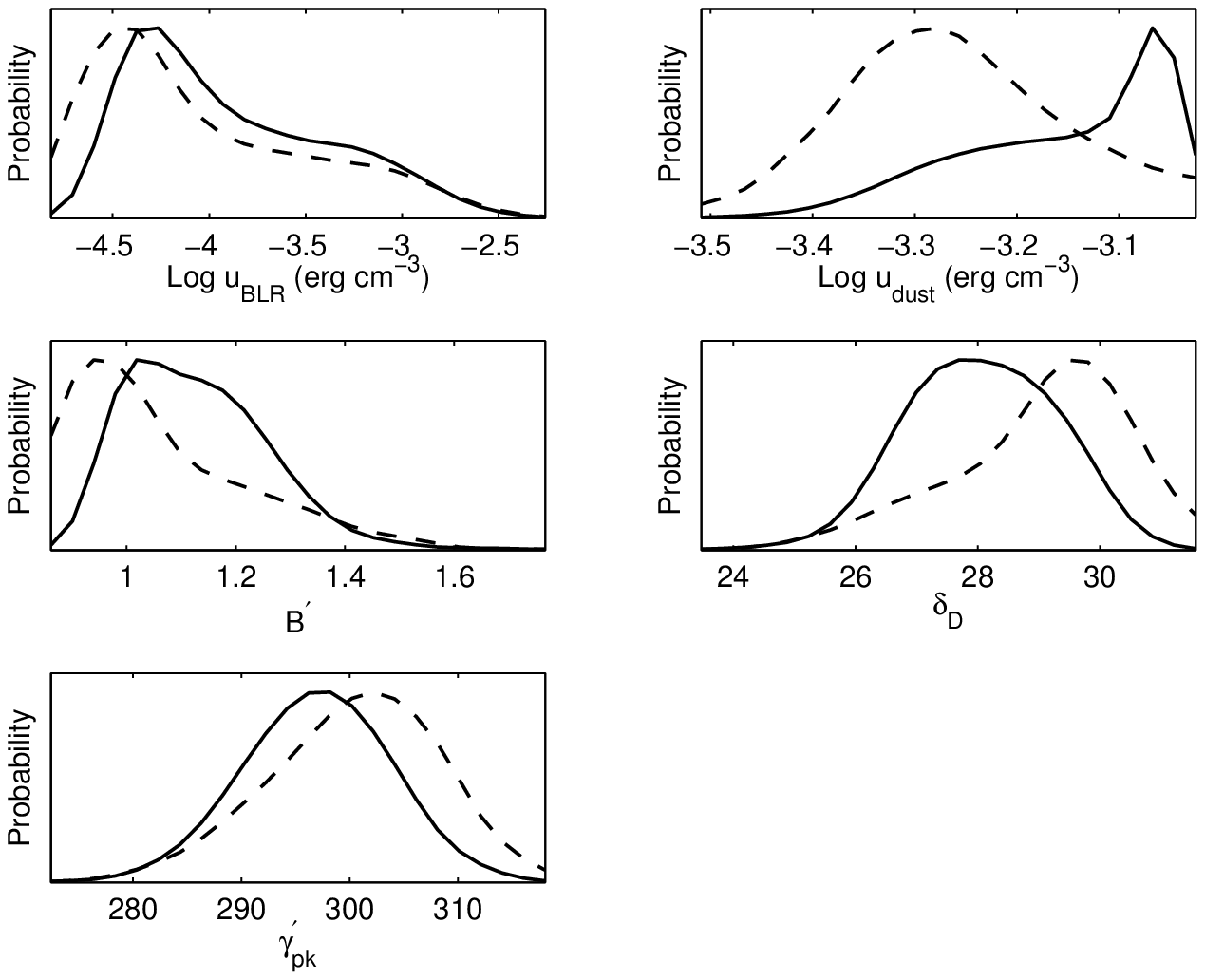}
		\includegraphics[width=240pt,height=180pt]{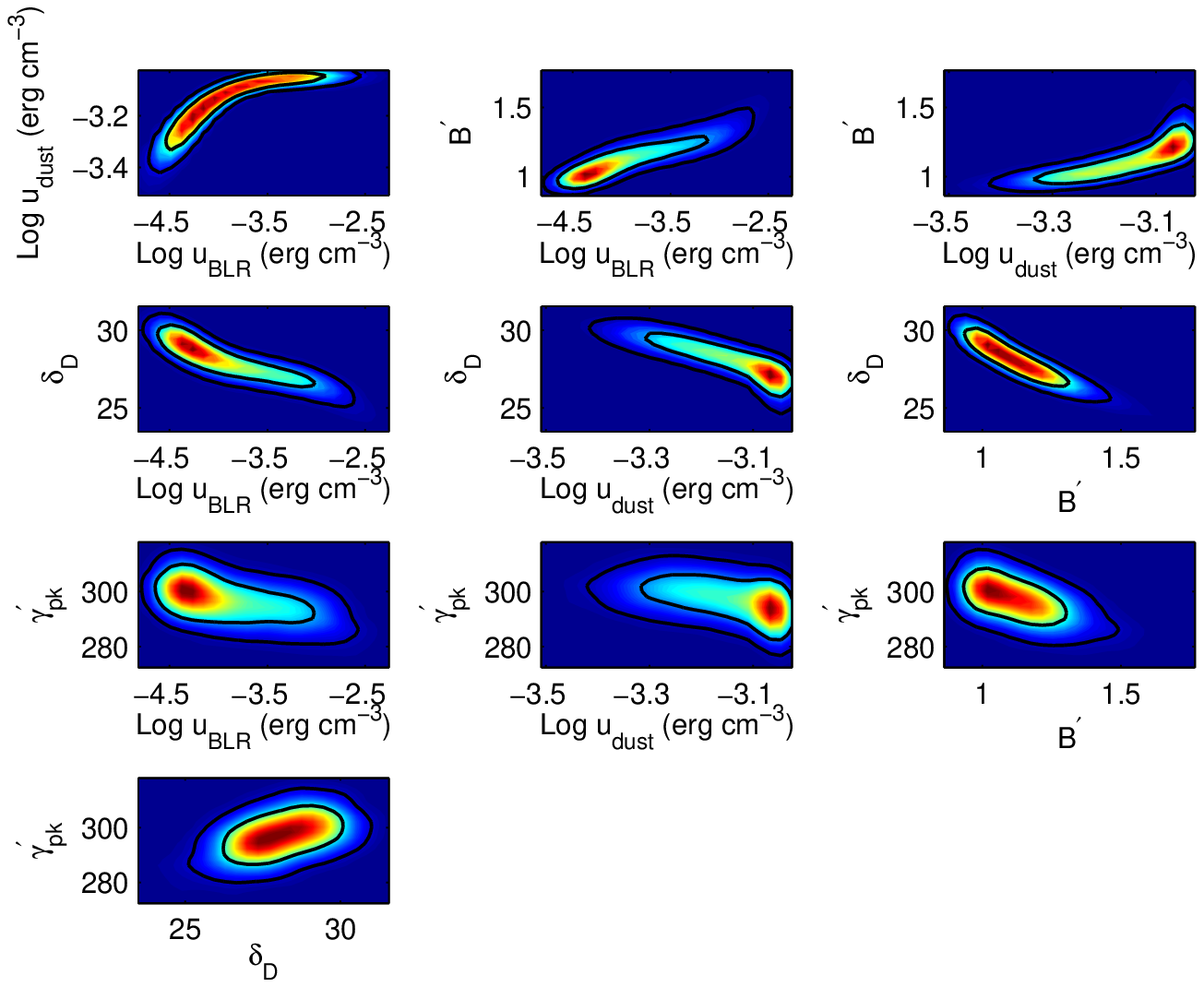}
	  \caption{One-dimensional probability distribution of parameter values (left; the dashed curves are the mean likelihoods of
samples and the solid curves are the marginalized probabilities) and two-dimensional contours of parameters (right; the regions enclosing 68\%(95\%) confidence level are shown) for Period C. The upper panel is for the input parameters, and the lower panel is for the output parameters. \label{distc}}
\end{figure*}

\section{Results}

We consider the SEDs of 3C 279 from simultaneous observations of \emph{Swift}, \emph{NuSTAR}, and \emph{Fermi}-LAT in two periods reported in \cite{Hayashida15}, namely Period A (2013 December 16-19) and Period C (2013 December 31-2014 January 3). Since an X-ray spectrum is lacking in Period B and there is no \emph{NuSTAR} data in Period D, we do not consider these SEDs. Note that the X-ray flux increased by a factor of $\approx 2$  and the $\gamma$-ray flux increased by a factor of $\approx 3$ from Period A to Period C.

The X-ray flux showed no intraday variations in Period A, but showed intraday variations in period C \citep{Hayashida15}. Therefore, the variability timescale for Period A is taken to be $t_4=5$ which corresponds to an observed variability timescale of $\sim$21 hours, and the variability timescale for Period C is set to be $t_4=1$ which corresponds to an observed variability timescale of $\sim$4 hours.

There is no evidence for a thermal emission feature in the two SEDs, making it difficult to constrain the accretion disk luminosity $L_{\rm disk}$.
We adopt $L_{\rm disk}=1.5\times10^{45}\rm \ erg s^{-1}$, which is the maximum disk luminosity allowed by the optical-UV SED.
Using this value, we derive $r_{\rm BLR}\simeq0.04\ $pc and $r_{\rm dust}\simeq0.4\ $pc.

By comparison, at a much earlier 3-week epoch between 1992 December and 1993 January,
\cite{1999ApJ...521..112P} originally discovered the accretion disk-emission from
3C 279 during a low-state monitored
with ROSAT, IUE, and EGRET, finding a  $\nu F_\nu$ flux of $\approx 3\times 10^{-12}$ erg s$^{-1}$ cm$^{-2}$
 between $\approx 1$ and
2$\times 10^{15}$ Hz,
corresponding to a UV luminosity $L_{\rm disk}=2.4\times10^{45}\rm \ erg s^{-1}$.

Due to the strong SSA below $10^{12}$ Hz, we fit only the optical, X-ray and $\gamma$-ray data,
but not the $230$ GHz SMA point.

Fig.~\ref{seda} shows the best-fit to the SED of 3C 279 for Period A. The fit is satisfactory.
The joint \emph{Swift} and \emph{NuSTAR} X-ray spectrum is successfully fit,
and is dominated by the SSC component.
The $\gamma$-ray emission is dominated by the EC-dust component.

In Fig.~\ref{dista}, we give the one-dimensional probability distributions and the two-dimensional contours of input and output parameters. The complete information on constraining parameters can be read from this figure.
We note that $\zeta_{\rm e}$ is constrained in the range of [0.8 - 1.7] (the marginalized 95\% confidence interval).
The distance $r$ is constrained in the range of [0.1 - 0.4] pc,
namely the emitting blob is outside the BLR, but is still inside the dust torus ($r_{\rm BLR}\simeq0.04\ $pc and $r_{\rm dust}\simeq0.4\ $pc). The derived $u_{\rm dust}$ is in the range of [5 - 9]$\times10^{-4}\rm \ erg\ cm^{-3}$, and $u_{\rm BLR}$ is in the range of [0.2 - 10]$\times10^{-4}\rm \ erg\ cm^{-3}$.
The magnetic field $B'$ is in the range of [0.7 - 1.2] G, and $\delta_{\rm D}$ is in the range of [17 - 20].
We summarise the marginalized 95\% confidence intervals for input and output parameters in Table~\ref{input}.
We also notice that there is a correlation between $r$ and $\zeta_{\rm e}$; and $B'$ is anti-correlated with $\delta_{\rm D}$ (see the two-dimensional contours).
These parameter correlations showed in Fig.~\ref{dista}  are caused by degeneracies of the leptonic model \citep[e.g.,][]{Tavecchio98,Sikora09,Dermer14}.
Nevertheless, all parameters except $u_{\rm BLR}$ are well constrained.

The fit to the SED from Period C is also satisfactory  (Fig.~\ref{sedc}). The \emph{Swift}-\emph{NuSTAR} X-ray spectrum from Period C shows a break at 3.7 keV with photon indices of 1.37 and 1.76, respectively, below and above the break energy, when fit by a broken power law \citep{Hayashida15}. Our results show that the one-zone leptonic model can successfully fit the X-ray spectrum. The X-ray emission in Period C is almost
entirely attributed to SSC, and the spectral break is naturally explained. The $\gamma$-ray emission is again dominated by the EC-dust component.

The one-dimensional probability distributions and the two-dimensional contours of input and output parameters derived in the fitting to the SED for Period C are shown in Fig.~\ref{distc}.
In this Period, the $\gamma$-ray emission site $r$ is in the range of [0.1 - 0.4] pc, identical to that in Period A;
but a larger $\zeta_{\rm e}$ of [3 - 5] and $\delta_{\rm D}$ of [26 - 30] is required to explain the higher $\gamma$-ray flux.
The magnetic field $B'$ of [0.9 - 1.4] is nearly identical to that in Period A.
From Period A to Period C, $\zeta_{\rm s}$ and $\nu_{14}$ increase, respectively, from $\sim$0.5 to 0.8, and from $\sim$0.3 to 0.5; and $b$ increases from $\sim$1.2 to 1.5.

In Table~\ref{input}, we also give the jet powers for a two-sided jet in the form of radiation ($P_{\rm r}$), Poynting flux ($P_{\rm B}$), where we assume that Doppler factor is related to the
bulk Lorentz factor through $\delta_{\rm D}=\Gamma$. The relativistic emitting electrons power $P_{\rm e}$, which is not shown, is obtained from the relation $P_{\rm e}$=$\zeta_{\rm e}P_{\rm B}$. This is the minimum jet power, and the addition of hadrons will only increase this power. One can see that the radiation power $P_{\rm r}$ is significantly greater than $P_{\rm B}$ and $P_{\rm e}$ in both cases.
\begin{table*}
\tiny
 \caption{Input and output parameters values. The mean values and the marginalized 95\% confidence intervals (CI) for interested parameters are reported. }
\label{input}
\begin{tabular}{@{}cccccccccccccccc}
 \hline
 & & & & & Input\\
 \hline
& $\zeta_e$ & $b$ & $L_{48}$ & $\nu_{14}$ & $t_4$ & $\zeta_s$ & $r$  & $T_{\rm dust}$ & $L_{\rm disk}$\\
&           &     &          &            &       &           & (pc) &  (K)           & ($\rm 10^{46}\ erg\ s^{-1}$)\\
\hline
Period A         & 1.12            & 1.24       & 0.18        & 0.30       &  5  & 0.47       & 0.2         & 800  & 0.15\\
 (95\% CI)    & 0.80-1.71       & 1.15-1.32  & 0.17-0.19   & 0.26-0.34  &  -  & 0.41-0.52  &0.1-0.4      &      &   -\\
 \hline
Period C         & 4.06            & 1.49       & 0.23        & 0.54       &  1  & 0.81       & 0.3         & 800  & 0.15\\
 (95\% CI)       & 3.04-5.0        & 1.43-1.57  & 0.22-0.24   & 0.50-0.58  &  -  & 0.74-0.89  &0.1-0.4      &      &   -\\
\hline
 & & & & & Output\\
 \hline
 & $B$ & $\delta_{\rm D}$ & $\gamma'_{\rm pk}$ & $u_{\rm dust}$ & $u_{\rm BLR}$ & $R'$             & $P_{\rm B}$               & $P_{\rm r}$ \\
 & (G) &                  &                    & ($10^{-3}\rm \ erg\ cm^{-3}$) & ($10^{-3}\rm \ erg\ cm^{-3}$) &($10^{16}\rm \ cm$) & ($\rm 10^{45}\ erg\ s^{-1}$)  & ($\rm 10^{46}\ erg\ s^{-1}$)\\
 \hline
Period A   & 1.0       & 18      & 260     &   0.7       &    0.2    & 2.7  &  1.7 & 2.0 \\
(95\% CI)  & 0.7-1.2   & 17-20   & 220-290 &   0.5- 0.9  &   0.02-1  &  -    &  -   & -\\
\hline
Period C   & 1.1       & 28      & 300     &   0.6       &    0.1    & 0.8  &  0.5 & 1.3 \\
(95\% CI)  & 0.9-1.4   & 26-30   & 280-310 &   0.5- 0.9  &   0.03-2  &  -    &  -   & -\\
\hline
\end{tabular}
\end{table*}

\section{Discussion and conclusions}

A one-zone leptonic model with a three-parameter log-parabola EED \citep{Dermer14} is used to fit
the two SEDs of 3C 279 where joint \emph{NuSTAR} and \emph{Swift} data are available \citep[the SEDs from Period A and Period C in][]{Hayashida15}. Using the MCMC method, we obtained the best-fit results and the uncertainties on the input and output parameters.
Our results show that the two SEDs can be successfully fit in near-equipartition conditions.
In both cases, the $\gamma$-ray emission is dominated by the EC-dust component, and the EC-BLR component is essentially negligible; and the X-ray emission is dominated by SSC emission. The SSC origin of X-ray emission is expected to be highly polarized if the magnetic field is perpendicular to the line of sight \citep[e.g.,][]{Krawczynski}.
The X-ray measurements of the degree of
polarization by future detectors such as
X-ray timing and polarization (XTP) and \emph{ASTRO-H}
would give more details on the magnetic field.

We take advantage of the MCMC technique to search the multi-dimensional parameter space, and to evaluate the uncertainties on the parameters. We have shown that all parameters except $u_{\rm BLR}$ are constrained very well by the current data. The stringently constrained parameters enable us to investigate the important issues in blazar physics confidently.

We derive the energy density of BLR radiation $u_{\rm BLR}<0.002\rm \ erg\ cm^{-3}$ at 95\% confidence level.
Our results show that the EC-BLR component is essentially
negligible in modelling the two SEDs 3C 279. Note that \citet{Cerruti} showed that EC-BLR component is necessary to explain the GeV break in 3C 454.3.
The distance $r$ from $\gamma$-ray emission blob to the black hole is well constrained in the range of [0.1 - 0.4] pc, namely the $\gamma$-ray emission region is outside the BLR, but is inside the dusty torus.
The value of $r$ depends on the assumptions on $r_{\rm BLR}$ and $r_{\rm dust}$.
Compared with the previous works \citep[e.g.,][]{ghisellini09,Hayashida,Nalewajko14,ghisellini14}, we adopted a smaller $r_{\rm dust}$. In Appendix~\ref{LargerIR}, we show the fitting results with $r_{\rm dust}=2\times10^{18}(L_{\rm disk}/10^{45}\rm \ erg\ s^{-1})^{1/2}\ $cm \citep{ghisellini14}. We find that the lower limit of $r$ derived in the fittings with the larger $r_{\rm dust}$ is still 0.1 pc; the upper limit is modified, but is still less than $r_{\rm dust}$. Therefore, our conclusion on the $\gamma$-ray emission site is still tenable. Our result is consistent with the $\gamma$-ray emission site in 3C 279 derived by \citet{Dermer14}, and is also consistent with the result derived by \citet{Nalewajko14} who used an independent method to constrain the $\gamma$-ray emission site.
\citet{Pacciani} explored the emission zone in high-energy flares of 10 FSRQs, and also found the evidence of gamma-ray flares occurring outside the BLR.

The log-parabola EED can be generated by stochastic particle acceleration, for example,
through systematic gyro-resonant acceleration of electrons
with plasma waves \citep[e.g.,][]{Becker06,Tramacere,Asano14,Kakuwa},
and the correlations between the curvature $b$ and the peak energy $\gamma'_{\rm pk}$ of EED provide evidence about the acceleration process \citep[e.g.,][]{Tramacere}. By fitting the SEDs, the EEDs are inferred, which helps to identify acceleration processes in the jet \citep{Yan13,Peng,Zhou14}. Indeed,
 \citet{Yan13} found that, because of the improved fits with
curving rather than power-law EEDs, it is likely that stochastic acceleration is
acting in the jet of Mrk 421 in the giant flare state in February 2010.
The acceleration process is also revealed by dynamic changes of the SED, which
implies changes in the EED. \citet{Yan13} also found that for Mrk 421 as $\nu_{14}$ increased from $\sim1000$ in the low state to $\sim10000$ in the giant flare, the curvature of PLLP EED in $N'(\gamma')$ distribution increased from $\sim1.7$ to $\sim3.8$.
We discussed that the changes in $\nu_{14}$ and the curvature in Mrk 421 are not
compatible with a purely acceleration-dominated scenario.
The two states of 3C 279 that we have analyzed indicates that a changing EED in terms of a changing log-parameter width parameter $b$ plays an important role in different spectral states.

\begin{figure}
	   \centering
		\includegraphics[width=250pt,height=190pt]{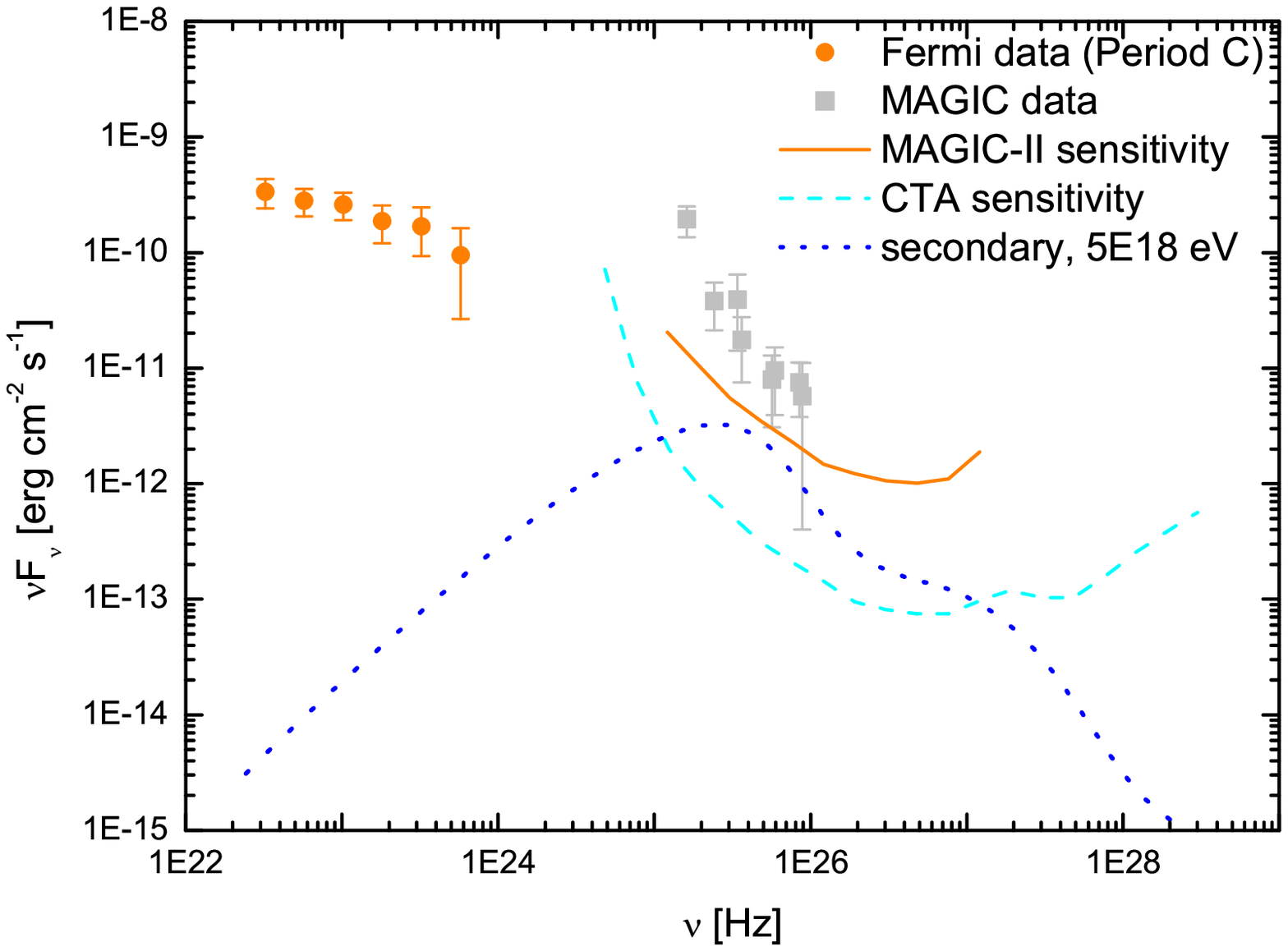}
	  \caption{Secondary $\gamma$-ray emission produced in the propagation of CRs. For comparison, the sensitivities of Imaging Atmospheric Cherenkov telescopes (MAGIC-II and CTA) and the non-simultaneous TeV MAGIC data for 3C 279 are shown.
The sensitivities and TeV data are obtained through the SED Builder of ASDC (http://tools.asdc.asi.it).   \label{CR}}
\end{figure}

It is known that X-ray spectra of several high-synchrotron-peaked (HSP) BL Lac objects can be described by a log-parabola function \citep[e.g.,][]{Massaro1,Massaro2}. Using a large data set of X-ray observations for several HSPs with $\nu_{14}>10$, \citet{Tramacere07,Tramacere09} and \citet{MassaroE} find that the synchrotron peak frequency is anti-correlated with the curvature parameter obtained by fitting the X-ray spectrum. \citet{Chen} finds that the synchrotron peak frequency is anti-correlated with the curvature parameter of the synchrotron spectrum for a sample of
 \emph{Fermi} blazar.

\citet{Tramacere} discussed the $\nu_{14}$ -- $b$ trends in an SSC model with stochastic acceleration.
They showed that the anti-correlation between $\nu_{14}$ and $b$ for HSPs could be explained in the stochastic acceleration SSC model by a change in the diffusion  process rather than by a change of magnetic field that affects
the cooling.
The radiative cooling of electrons in FSRQs is more efficient than that in HSPs,
which may lead to the different $\nu_{14}$ -- $b$ trend from that for HSPs.
Our results show that both $\nu_{14}$ and $b$ significantly increase from period A to period C. More high-quality SEDs are needed to confirm the $\nu_{14}$ -- $b$ trend for FSRQs.
Moreover, our results indicate that the $\gamma$-ray emission region is in the dust torus, in
contrast to \citet{zhang13}, who assumed a absence of the dust torus.
 The unknown $\gamma$-ray emission site may complicate the $\nu_{14}$ -- $b$ trend for FSRQs.
Because of the larger photon energy density in the
BLR, radiative cooling of electrons is stronger than in the dust torus.
On the other hand, the relatively inefficient radiative cooling of electrons in the dust tours allows the electrons to be accelerated to higher energies by the inefficient stochastic mechanism.

\begin{figure}
	   \centering
		\includegraphics[width=250pt,height=190pt]{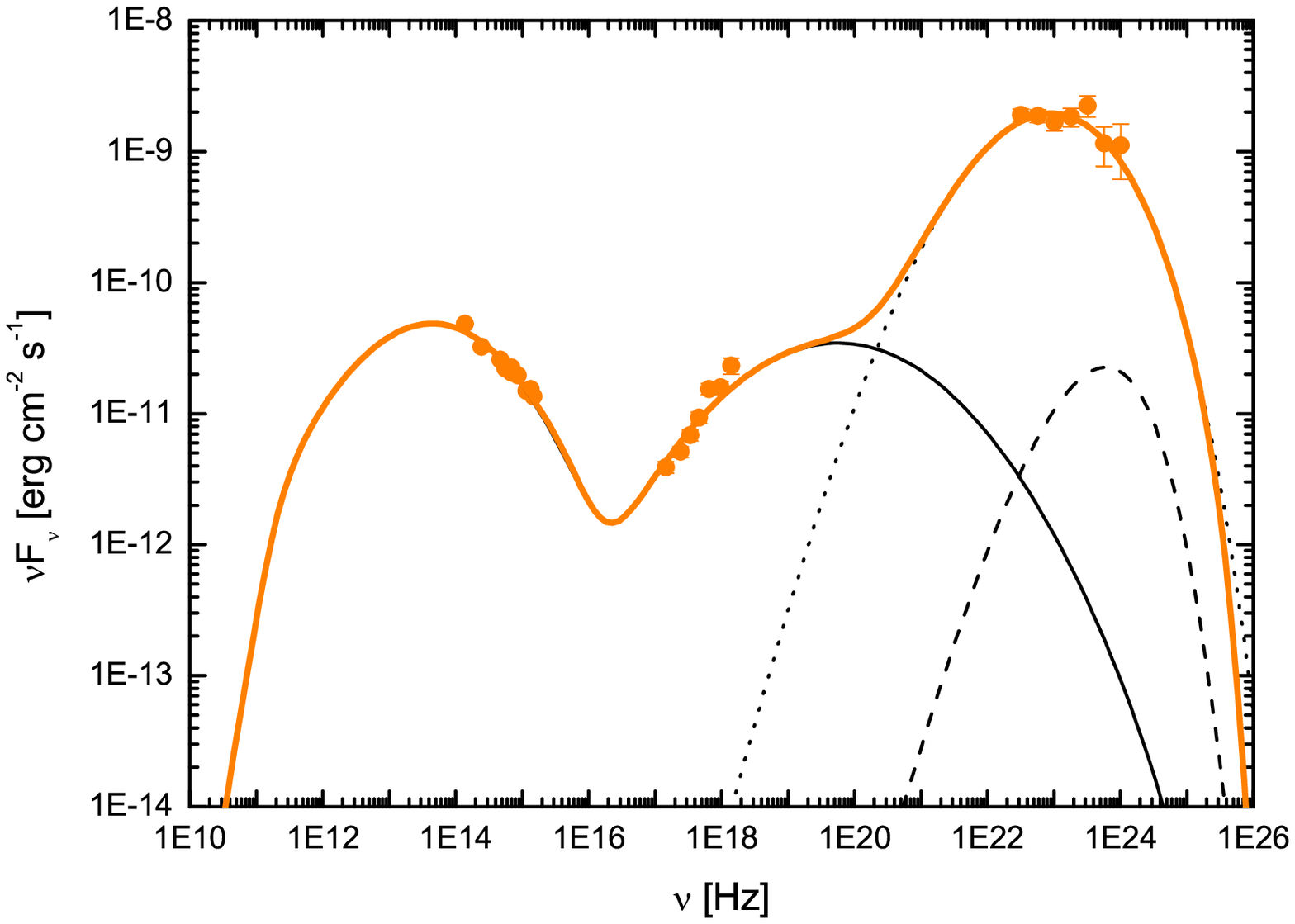}
	  \caption{Best-fit model to the SED of 3C 279 \citep{Hayashida15} for Period D. The parameter values are $u_{\rm dust}\sim0.6\times10^{-3}\rm \ erg\ cm^{-3}$, $\zeta_{\rm s}\sim1.0$, $\nu_{14}\sim0.9$, $b\sim1.4$ and $L_{48}\sim0.3$. \label{pd}}
\end{figure}

In our model, the radiation power is much greater than the magnetic-field and relativistic electron power, assuming
that the baryon-loading (the ratio of the energy density of hadrons to electrons, $\eta_{\rm bl}$) is low,
and will not have an impact on the total jet power ($P_{\rm r}$+$P_{\rm B}$+$P_{\rm e}$+$P_{\rm h}$, where $P_{\rm h}$ is the power carried by hadrons) as long as $\eta_{\rm bl}\lesssim 10$. Assuming $\eta_{\rm bl}=1$, we derive $P_{\rm h}=\eta_{\rm bl}P_{\rm e}=0.3\times10^{46}\rm \ erg\ s^{-1}$ and $0.2\times10^{46}\rm \ erg\ s^{-1}$ for Period A and Period C, respectively. 3C 279 has a black hole with mass (3-8)$\times10^8M_\odot$ \citep[e.g.,][]{Gu,Woo}. The Eddington luminosity is therefore in the range (4-10)$\times10^{46}\rm \ erg\ s^{-1}$.
For low baryon loading, the total jet powers in our model are comfortably below the Eddington luminosity of 3C 279.

Blazars are usually considered as cosmic-ray (CR) sources. Using the near-equipartition-log-parabola model, \citet{Dermer15} gave the following expression for maximum escaping proton energy
\begin{equation}
E_{\rm max}({\rm eV}) = 1.4\times 10^{20}  L_{48}^{5/16}\;({t_4\over \nu_{14}})^{1/8} {f_1^{1/4} f_2^{1/8}\over \zeta_e^{1/4} \zeta_s^{1/16}f_0^{11/16}}\;,
\label{Emax}
\end{equation}
with $f_0=1/3$, $f_1=10^{-1/4b}$, and $f_2=10^{1/b}$ .
Using the best-fit values for Period C (Table~\ref{input}), we derive $E_{\rm max}\sim10^{20} - 2\times10^{20}\ $eV.
However, it should be noted that 3C 279 cannot contribute to the CRs with energy $\gtrsim10^{18}\ $eV at the Earth, because of the significant energy losses during the CRs traveling to us.

Besides $E_{\rm max}$, the other key quantity of a blazar is its CR luminosity $L_{\rm CR}$. Secondary $\gamma$ rays are produced when high-energy CRs propagate towards us \citep[e.g.,][]{Kalashev09,Kalashev13}, which depends primarily on the values of $L_{\rm CR}$ and $E_{\rm max}$. Such kinds of non-variable secondary emission \citep[variable on timescale of years;][]{Prosekin} have been proposed to explain the steady VHE ($\gtrsim 100$ GeV)  emission from distant blazars \citep[e.g.,][]{essey10b,essey11,murase12,Aharonian2013,Takami13,Yan15b}.

If 3C 279 emits high-energy cosmic rays, the secondary emission should be lower than its VHE emission, which puts an upper limit on $L_{\rm CR}$. There are no recently reported VHE observations for 3C 279, so we use the sensitivity of MAGIC-II as the upper limit of secondary emission. In Fig.~\ref{CR}, we show the secondary gamma-ray spectrum calculated by using the code \texttt{TransportCR} \citep{Arisaka,Gelmini12,Kalashev14} and the EBL model of \cite{Franceschini} \citep[see][for detailed comparisons for various EBL models]{Finke10}. In the calculation, we assume a log-parabola CR distribution with $b=1.48$ and peak energy $E_{\rm pk}=5\times10^{18}\ $ eV, and let the intergalactic magnetic field strength $B_{\rm IGMF}=10^{-15}\ $G \citep[e.g.,][]{essey-AK11} and its coherence length $\lambda_{\rm coh}=1\ $Mpc.
To make the secondary emission lower than the sensitivity of MAGIC-II, the CR luminosity is required to be $<6\times10^{46}\rm \ erg\ s^{-1}$, significantly lower than $L_{\rm syn}$ and the apparent gamma-ray luminosity ($\sim10^{48}\rm \ erg\ s^{-1}$).
The upper limit of the corresponding absolute CR power $L^+_{\rm ab,CR}$ is $\sim10^{44}\rm \ erg\ s^{-1}$ (adopting $\delta_{\rm D}=28$ in Period C), a quarter of $P_{\rm B}$ ($P_{\rm B}$ is for a two-sided jet) in Period C. In Fig.~\ref{CR}, it can be seen that Cherenkov Telescope Array (CTA), having significantly improved sensitivity over the present imaging air Cherenkov arrays, will put stronger constraints on the CR luminosity. The upper limit assumes that ultra-high energy cosmic rays are not deflected and isotropized during transit, which is unlikely unless the blazar is found in structured regions, for example, clusters and filaments \citep{murase12}. Even if the UHECRs can freely escape from the source region, they can be deflected during transport across intergalactic space \citep{takami15}.

If the hadrons in the jet are relativistic cosmic rays, as discussed above, these protons could,
before escaping from the emission region,
lose energies and produce secondary radiations via synchrotron and photohadronic
interactions with low-energy photons \citep[e.g.,][]{Dermer142,murase14}.
However, the emission made by the protons in the emission region is negligible
compared to the observed data, because the proton power ($P_{\rm h}$) in our model is 2 -- 3 orders of magnitude below the proton power required by hadronic model interpretations of FSRQ SEDs ($10^{47}$-$10^{49}\rm \ erg\ s^{-1}$) and the magnetic field in our model is 1 -- 2 orders of magnitude below that usually
found in hadronic models (10-100 G) \citep{bottcher13,Diltz15}.

In this paper, we fitted the SEDs for Period A and Period C reported by \citet{Hayashida15} with our one-zone leptonic model.
This model can also explain the SED in the bright flare state of Period D with $F(>100\rm\ MeV)$ of $10^{-5}\rm \ photons\ cm^{-2}\ s^{-1}$ and $t_{\rm obs,var}\sim2\ $hours reported by \citet{Hayashida15}. In Fig.~\ref{pd}, we show our best-fit result for the SED of Period D in \citet{Hayashida15}.
In this fit, we use $t_4=0.5$ and $T_{\rm dust}=800\ $K. The best-fit model requires $\zeta_{\rm e}\sim10$ and $r\sim0.2\ $pc.
The deduced parameters are $B\sim0.8\ $G, $\delta_{\rm D}\sim42$, $\gamma'_{\rm pk}\sim360$ and $R'\sim6\times10^{15}\ $cm (see the other parameters values in the caption in Fig.~\ref{pd}).
The results of Period D, compared with those of Periods A and C,
 are consistent with $b$ and $\nu_{14}$ mutually increasing.
In contrast, the HSP BL Lacs with large peak synchrotron
frequencies have broader widths of the synchrotron SEDs (corresponding
to smaller values of $b$) than FSRQs with smaller values of
$\nu_s$ \citep{Chen}.

Because of the lack of X-ray spectrum and the poor data coverage of the synchrotron hump in Period B,
we do not fit the SED in Period B.
We found that the EC-dust emission in our model could account for the very hard $\gamma$-ray spectrum ($\Gamma_{\gamma}\simeq1.7$) in Period B with $\nu_{14}\sim2.6$, $L_{48}\sim0.07$, $b\sim1.7$, $\zeta_{\rm e}\sim3$, $t_4\sim0.5$ and $r\sim0.3\ $pc; and we derived $B\sim0.3\ $G, $\delta_{\rm D}\sim60$, $\gamma'_{\rm pk}\sim950$, and $u_{\rm dust}\sim0.6\times10^{-3}\rm erg\ cm^{-3}$.

In conclusion, we have applied the MCMC technique developed for blazar studies
by \citet{Yan13} to the high-quality simultaneous data of 3C 279 \citep{Hayashida15},
using a log-parabola EED \citep{Dermer14}.  We find that the $\gamma$-ray SED is dominated by the
dusty torus radiation, whereas the contribution of the BLR radiation
field is very weak. This allows us to place the $\gamma$-ray emission
region outside the BLR, at $\gtrsim 0.1$ pc, but within the IR
radiation field  of the torus. The reasonable fits obtained
with curving particle distributions are more simply expained
with a stochastic acceleration mechanism.  The allowed parameter range
in 3C 279 is well
constrained at different epochs, and shows a positive correlation of the log-parabola
width parameter $b$ with peak synchrotron frequency $\nu_s$. Analysis of more FSRQ SEDs will be required to
determine if this correlation is robust.

\section*{Acknowledgments}
 We would like to give special thanks to Charles Dermer, who carefully edited the manuscript and gave us many valuable suggestions and comments.  We thank Charles Dermer, J.\ Finke, Kinwah Wu for discussions, and the anonymous
referee for very helpful and constructive questions.
 We are grateful to Krzysztof Nalewajko for providing us the data of 3C 279. This work is partially supported by the National Natural Science Foundation of China (NSFC 11433004) and Top Talents Program of Yunnan Province, China. DHY acknowledges partial funding support by China Postdoctoral Science Foundation under grant No. 2015M570152.
 SNZ acknowledges partial funding support by 973 Program of China under grant 2014CB845802, by the National Natural Science Foundation of China (NSFC) under grant Nos. 11133002 and 11373036, by the Qianren start-up grant 292012312D1117210, and by the Strategic Priority Research Program ``The Emergence of Cosmological Structures'' of the Chinese Academy of Sciences (CAS) under grant No. XDB09000000. We acknowledge the use of the ASDC web Tools (SED Builder, http://tools.asdc.asi.it).

\bibliography{refernces}


\appendix
\section{Fitting results with larger size of dust torus}
\label{LargerIR}
We show the fitting results with a larger $r_{\rm dust}$, i.e., $r_{\rm dust}=2\times10^{18}(L_{\rm disk}/10^{45}\rm \ erg\ s^{-1})^{1/2}\ $cm \citep[e.g.,][]{ghisellini14}. We then have $r_{\rm dust}\simeq0.8\ $pc.
Using $\tau_{\rm dust}=0.3$, we derive $u_{\rm dust}\lesssim2\times10^{-4}\rm \ erg\ cm^{-3}$.

\begin{figure}
	   \centering
		\includegraphics[width=240pt,height=180pt]{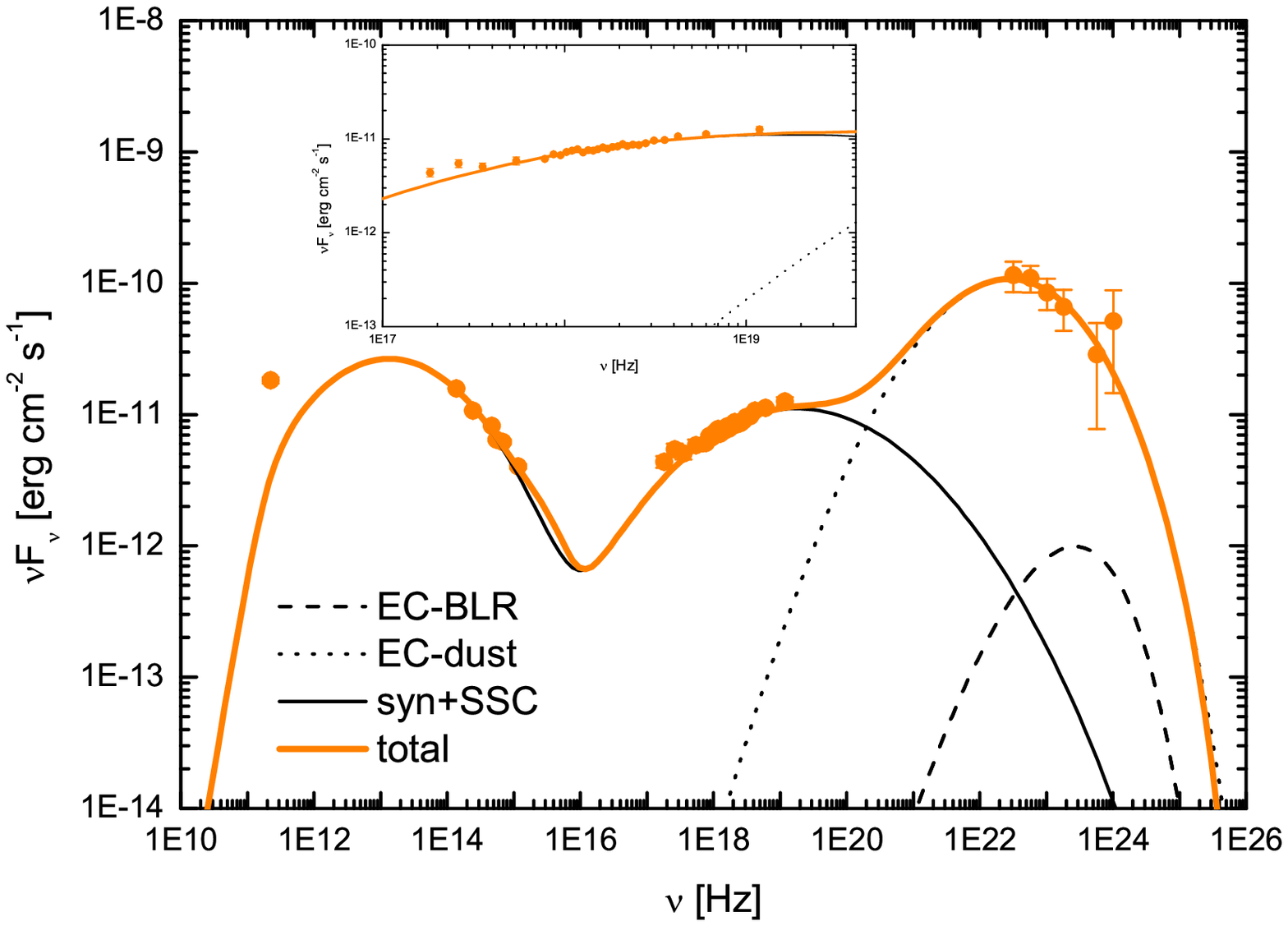}
	 	\includegraphics[width=240pt,height=180pt]{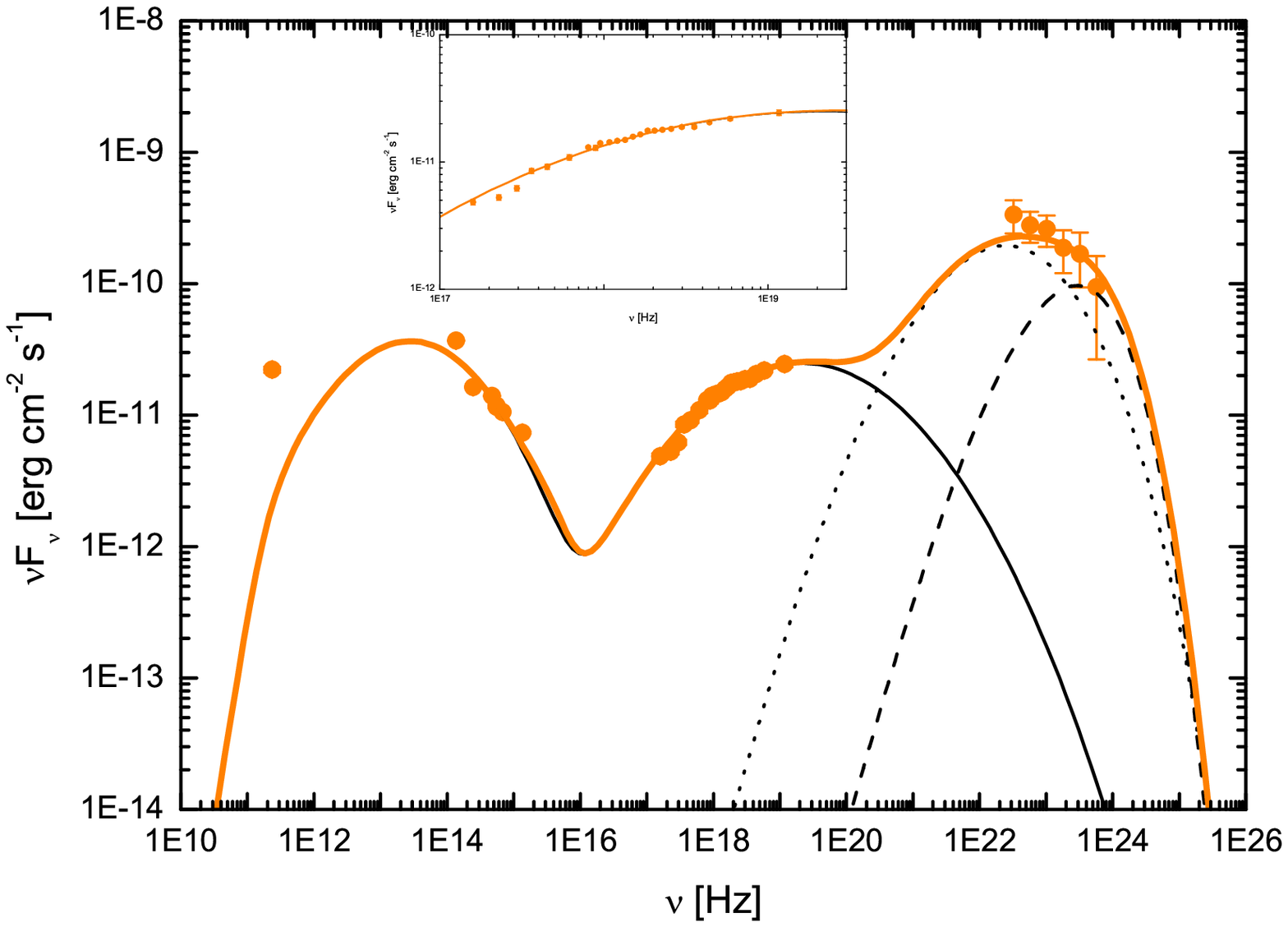}
	  \caption{Best-fit models with a larger $r_{\rm dust}$  to the SEDs of 3C 279. \emph{Upper}: for Period A; \emph{lower}: for Period C. The inset shows the details of fit at X-ray energies. \label{sedac}}
\end{figure}

Figure \ref{sedac} shows the best-fit results with this larger $r_{\rm dust}$ .
One can see that the fit to the SED in Period A is comparable to the fit showed in Section 3;
however, the fit to lowest $\gamma$-ray data in Period C is bad in the case of the larger $r_{\rm dust}$. 

\begin{figure}
	   \centering
		\includegraphics[width=240pt,height=180pt]{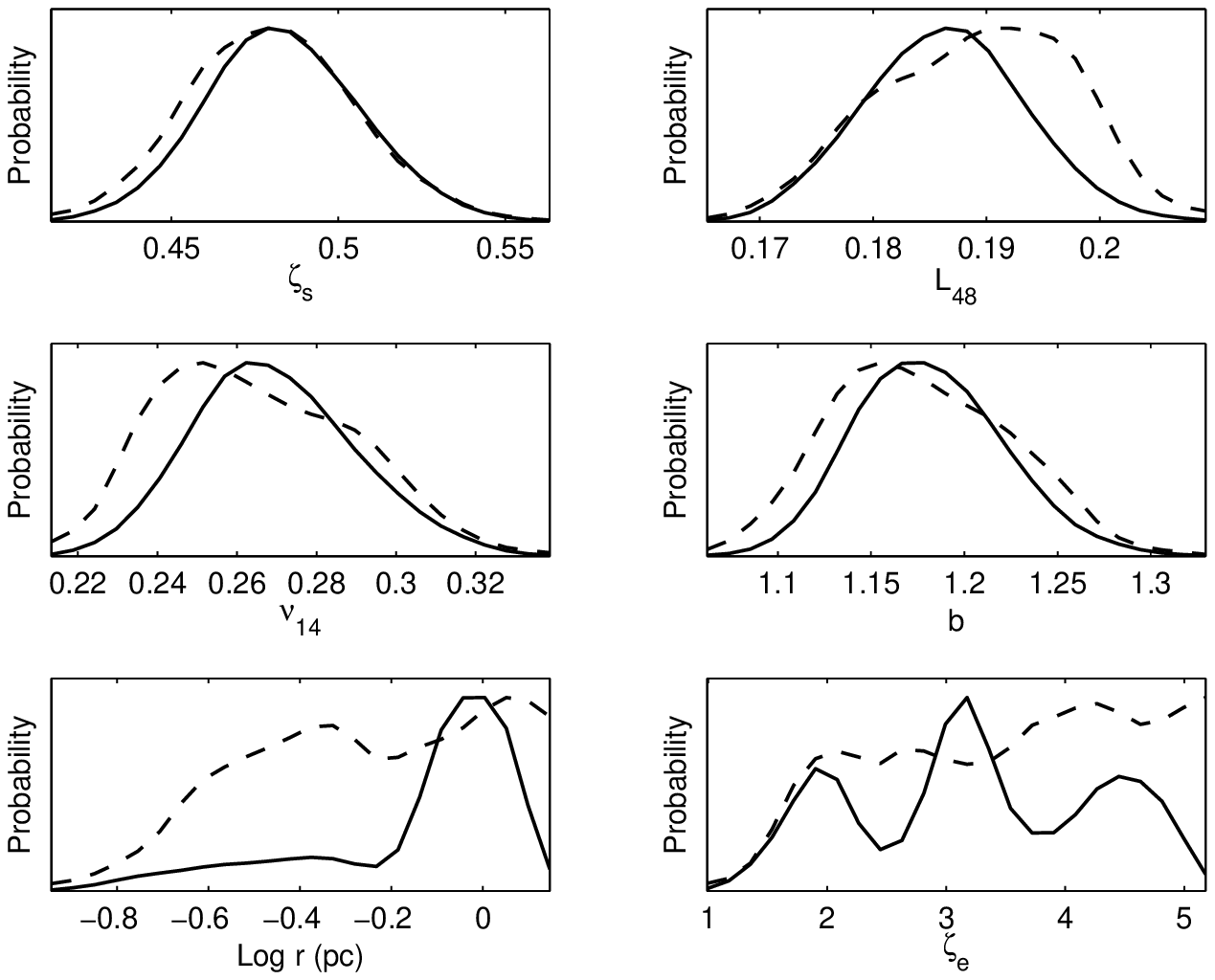}
		\includegraphics[width=240pt,height=180pt]{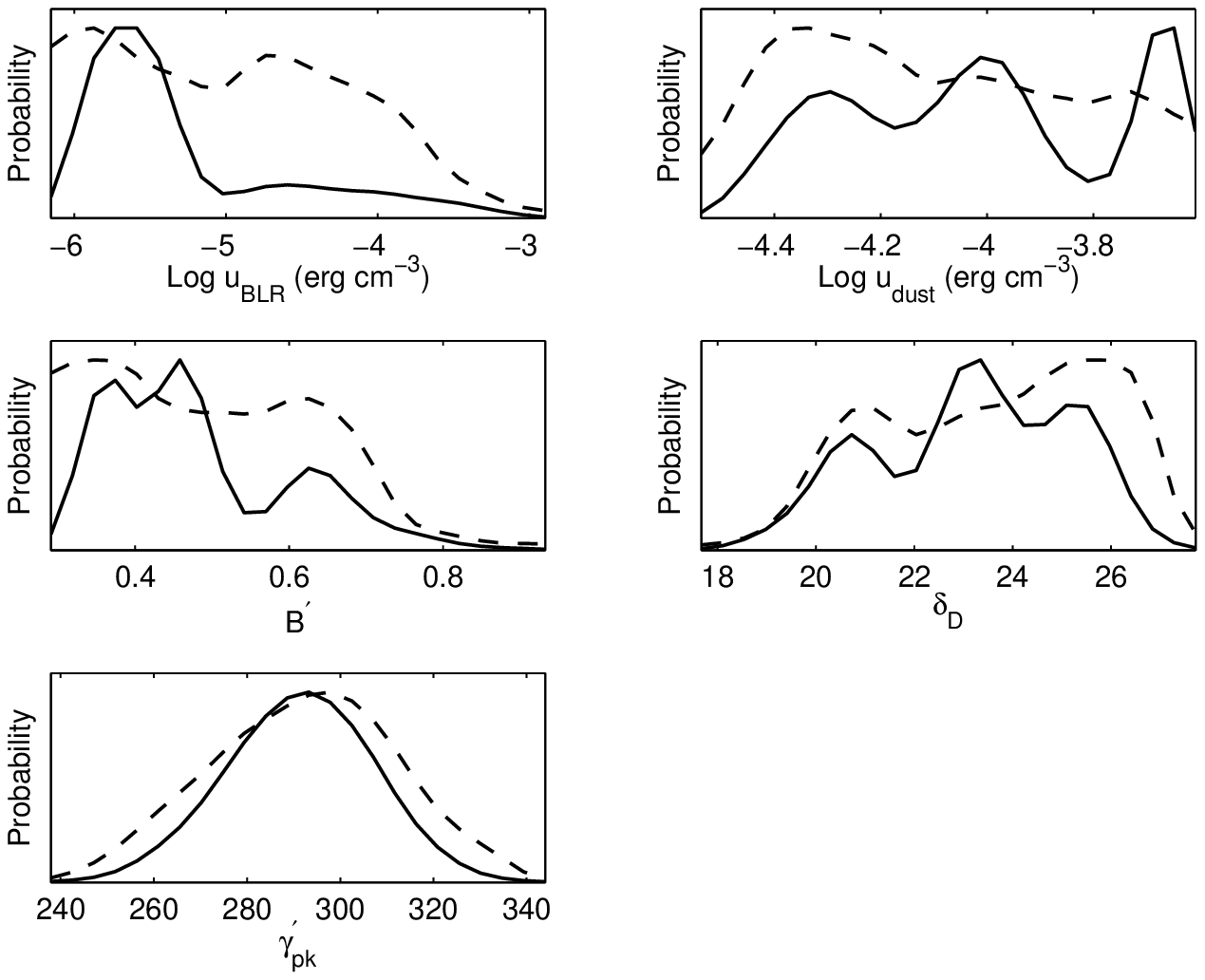}
	  \caption{One-dimensional probability distribution of parameter values for Period A. \label{dista1}}
\end{figure}

\begin{figure}
	   \centering
		\includegraphics[width=240pt,height=180pt]{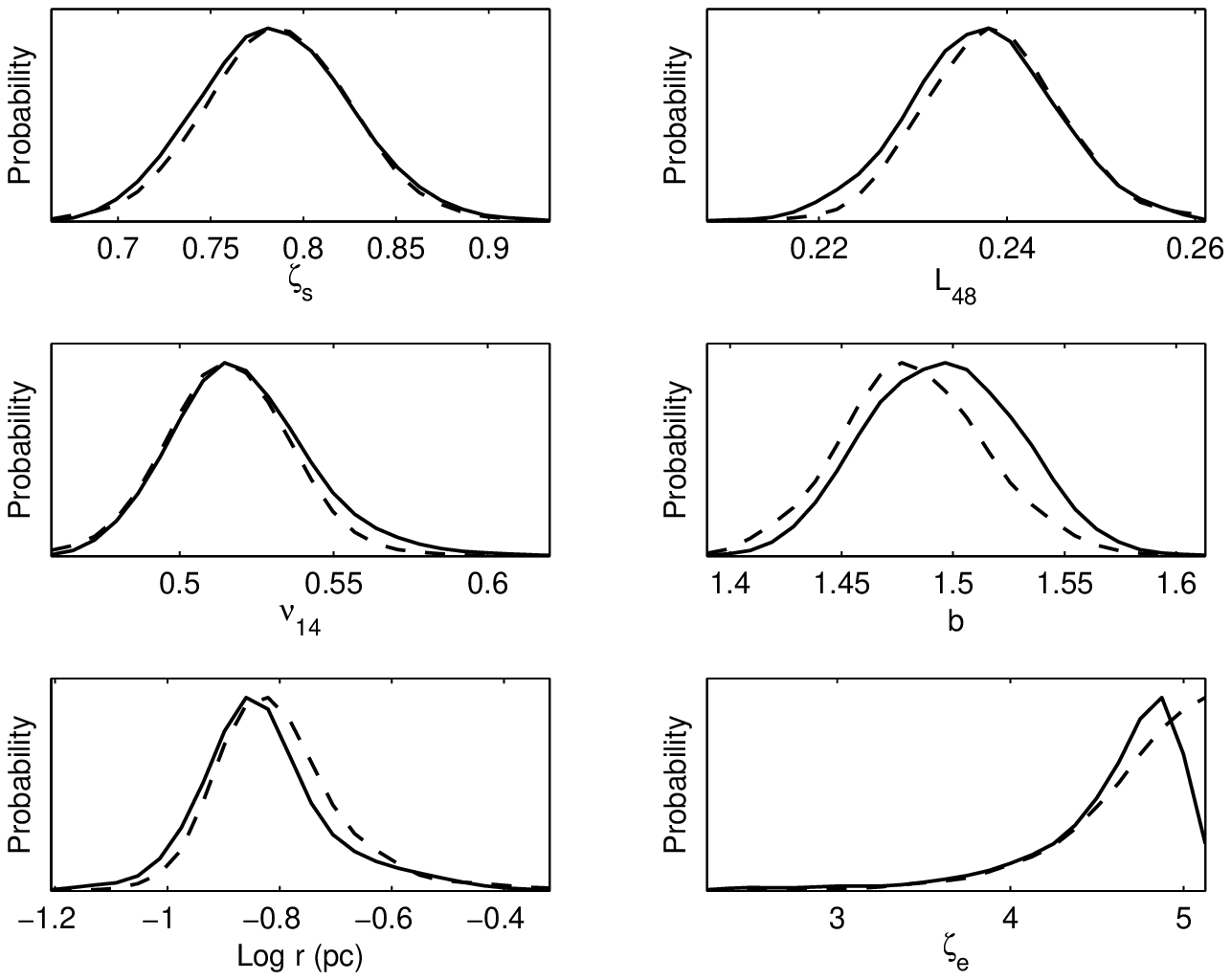}
		\includegraphics[width=240pt,height=180pt]{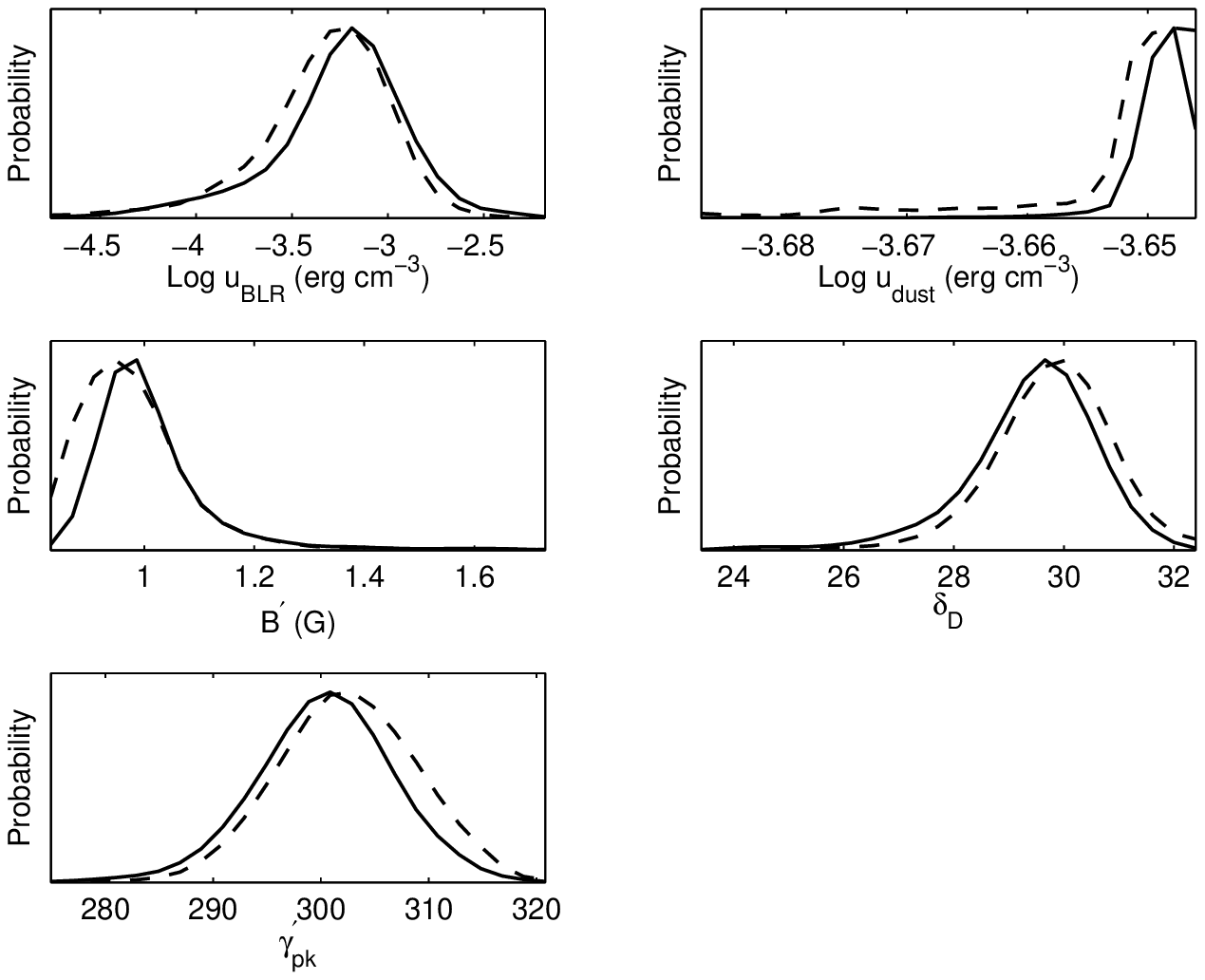}
	  \caption{One-dimensional probability distribution of parameter values for Period C. \label{distc1}}
\end{figure}

In Figs.~\ref{dista1} and \ref{distc1}, we show the one-dimensional probability distribution of parameter values.
The mean values and the marginalized 95\% confidence intervals of the model parameters are reported in Table~\ref{input1}.
In Period A, a larger $\zeta_{\rm e}$ and $r$ is required in fitting with the larger $r_{\rm dust}$.
The other input parameters change little, compared with those given in Section 3. The change in $\zeta_{\rm e}$ leads to the significant changes in $B'$ and $\delta_{\rm D}$.
In Period C, the input parameters are nearly identical to those reported in Section 3.

\begin{table*}
\tiny
 \caption{Input and output parameters values derived in the fittings with the larger $r_{\rm dust}$. The mean values and the marginalized 95\% confidence intervals (CI) for interested parameters are reported. }
\label{input1}
\begin{tabular}{@{}cccccccccccccccc}
 \hline
 & & & & & Input\\
 \hline
& $\zeta_e$ & $b$ & $L_{48}$ & $\nu_{14}$ & $t_4$ & $\zeta_s$ & $r$  & $T_{\rm dust}$ & $L_{\rm disk}$\\
&           &     &          &            &       &           & (pc) &  (K)           & ($\rm 10^{46}\ erg\ s^{-1}$)\\
\hline
Period A         & 3.25            & 1.18       & 0.19        & 0.27       &  5  & 0.48       & 0.7         & 800  & 0.15\\
 (95\% CI)       & 1.52-4.91       & 1.11-1.26  & 0.17-0.20   & 0.23-0.31  &  -  & 0.44-0.53  &0.1-1.2      &      &   -\\
 \hline
Period C         & 4.58            & 1.50       & 0.24        & 0.52       &  1  & 0.79       & 0.2         & 800  & 0.15\\
 (95\% CI)       & 3.13-4.98       & 1.43-1.56  & 0.22-0.25   & 0.48-0.57  &  -  & 0.71-0.87  &0.1-0.3      &      &   -\\
\hline
 & & & & & Output\\
 \hline
 & $B$ & $\delta_{\rm D}$ & $\gamma'_{\rm pk}$ & $u_{\rm dust}$ & $u_{\rm BLR}$ & $R'$             & $P_{\rm B}$               & $P_{\rm r}$ \\
 & (G) &                  &                    & ($10^{-3}\rm \ erg\ cm^{-3}$) & ($10^{-3}\rm \ erg\ cm^{-3}$) &($10^{16}\rm \ cm$) & ($\rm 10^{45}\ erg\ s^{-1}$)  & ($\rm 10^{46}\ erg\ s^{-1}$)\\
 \hline
Period A   & 0.5       & 23      & 290     &   0.1         &    0.005      &3  &  0.9 & 0.6 \\
(95\% CI)  & 0.3-0.8   & 19-26   & 260-320 &   0.04- 0.25  &    0.001-0.3  &  -    &  -   & -\\
\hline
Period C   & 1.0       & 29      & 300     &   0.21         &    0.6      & 0.8  &  0.5 & 1.0 \\
(95\% CI)  & 0.9-1.3   & 27-31   & 290-310 &   0.20- 0.22   &   0.07-2  &  -    &  -   & -\\
\hline
\end{tabular}
\end{table*}

\bsp	
\label{lastpage}
\end{document}